\documentclass[journal]{IEEEtran}

\usepackage{geometry} 
\usepackage{placeins}
\usepackage{pifont}
\usepackage[utf8]{inputenc}
\usepackage{hyperref}
\usepackage{amsmath,amsfonts,amsthm} 
\usepackage{amssymb,mathtools}
\usepackage[dvipsnames]{xcolor} 
\usepackage{graphicx,wrapfig,caption,subcaption} 
\usepackage{insdljs,lcg} 
\usepackage{framed,adjustbox}
\usepackage[most]{tcolorbox} 
\usepackage{booktabs}
\usepackage{cite}
\usepackage{tabu,multirow}
\usepackage{boldline,blindtext}
\usepackage[super]{nth} 
\usepackage[official]{eurosym} 
\usepackage{fontawesome} 
\usepackage{algpseudocode}
\usepackage{algorithm}
\usepackage{pgfplots}
\usepackage{colortbl}
\usepackage{sansmath}

\usepgfplotslibrary{groupplots}
\pgfplotsset{compat=newest}

\pgfplotsset{every axis/.append style={
		label style={font=\footnotesize},
		tick label style={font=\footnotesize}  
}}

\usetikzlibrary{decorations.markings}
\usetikzlibrary{datavisualization.formats.functions,arrows.meta}

\tikzstyle{sum} = [draw, fill=blue!20, circle, node distance=1cm]
\tikzstyle{input} = [coordinate]
\tikzstyle{output} = [coordinate]
\tikzstyle{pinstyle} = [pin edge={to-,thin,black}]

\pgfdeclarelayer{background}
\pgfdeclarelayer{foreground}
\pgfsetlayers{background,main,foreground}

\tikzstyle{block}=[draw, thick, fill = White, text width=2.5em, text centered, minimum height=2em]
\tikzstyle{sumprod} = [draw,thick,circle,fill=White]
\tikzstyle{ann} = [above, text width=5em]
\def\blockdist{1}
\def\edgedist{1.9}

\definecolor{GU}{RGB}{255,0,0}
\definecolor{GR}{RGB}{237,177,32}
\definecolor{GS}{RGB}{119,172,48}
\definecolor{GT}{RGB}{77,190,238}

\newcommand{\kin}{_{\mathrm{kin}}}

\DeclareMathOperator{\argmin}{argmin}

\newcommand{\Jcal}{\mathcal{J}}

\newcommand{\Fxf}{F_{x\rm{f}}}
\newcommand{\Fxr}{F_{x\rm{r}}}
\newcommand{\Fyf}{F_{y\rm{f}}}
\newcommand{\Fyr}{F_{y\rm{r}}}
\newcommand{\Fzf}{F_{z\rm{f}}}
\newcommand{\Fzr}{F_{z\rm{r}}}

\newcommand{\alf}{\alpha_{\rm{f}}}
\newcommand{\alr}{\alpha_{\rm{r}}}

\newcommand{\dltf}{\delta_{\rm{f}}}

\newcommand{\lf}{l_{\rm{f}}}
\newcommand{\lr}{l_{\rm{r}}}
\newcommand{\e}{_{\rm{ego}}}
\newcommand{\obs}{_{\rm{obs}}}
\newcommand{\rf}{_{\mathrm{ref}}}

\newcommand{\Np}{N_{\mathrm{p}}}

\newcommand{\ts}{t_{\rm{s}}}

\newcommand{\barr}{_{\rm{safe}}}
\newcommand{\vel}{_{\rm{brake}}}
\newcommand{\steer}{_{\rm{steer}}}
\newcommand{\stab}{_{\rm{stable}}}
\newcommand{\des}{_{\rm{des}}}

\newcommand{\mpc}{_{\rm{MPC}}}
\newcommand{\ff}{_{\rm{MSF}}}

\theoremstyle{definition}

\newtheorem{exmp}{Example}
\newtheorem{rem}{Remark} 
\newtheorem{obsrv}{Observation}

\usetikzlibrary{positioning}
\usetikzlibrary{arrows.meta}

\ifCLASSINFOpdf
\else
\fi
\begin{document}

\title{Dodging the Moose: Experimental Insights in Real-Life Automated Collision Avoidance}

\author{
Leila Gharavi,
Simone Baldi,~\IEEEmembership{Senior Member,~IEEE},
Yuki Hosomi,~\IEEEmembership{Student Member,~IEEE},
Tona Sato,~\IEEEmembership{Student Member,~IEEE},
Bart De~Schutter,~\IEEEmembership{Fellow,~IEEE},
Binh-Minh Nguyen,~\IEEEmembership{Member,~IEEE},
Hiroshi Fujimoto,~\IEEEmembership{Fellow,~IEEE}
\thanks{Leila Gharavi and Bart De~Schutter are with Delft Center for Systems and Control, Delft University of Technology, Delft, the Netherlands.}
\thanks{Simone Baldi is with the School of Mathematics, Southeast University, Nanjing, China.}
\thanks{Yuki Hosomi, Tona Sato, Binh-Minh Nguyen and Hiroshi Fujimoto are with the Department of Advanced Energy, Graduate School of Frontier Sciences, The University of Tokyo, Chiba 277-8561, Japan.}
\thanks{Experiment videos are accessible on YouTube: \url{https://youtu.be/GQk1n9bXKJU?si=IMxbQAHhFramVYBR}}
}

\markboth{Draft for Journal Submission}%
{Shell \MakeLowercase{\textit{et al.}}: Bare Demo of IEEEtran.cls for IEEE Journals}



\maketitle

\begin{abstract}
The sudden appearance of a static obstacle on the road, i.e.~the moose test, is a well-known emergency scenario in collision avoidance for automated driving. Model Predictive Control (MPC) has long been employed for planning and control of automated vehicles in the state of the art. However, real-time implementation of automated collision avoidance in emergency scenarios such as the moose test remains unaddressed due to the high computational demand of MPC for evasive action in such hazardous scenarios. This paper offers new insights into real-time collision avoidance via the experimental implementation of MPC for motion planning after a sudden and unexpected appearance of a static obstacle. As the state-of-the-art nonlinear MPC shows limited capability to provide an acceptable solution in real-time, we propose a human-like feed-forward planner to assist when the MPC optimization problem is either infeasible or unable to find a suitable solution due to the poor quality of its initial guess. We introduce the concept of maximum steering maneuver to design the feed-forward planner and mimic a human-like reaction after detecting the static obstacle on the road. Real-life experiments are conducted across various speeds and level of emergency using FPEV2-Kanon electric vehicle. Moreover, we demonstrate the effectiveness of our planning strategy via comparison with the state-of-the-art MPC motion planner.
\end{abstract}


\IEEEpeerreviewmaketitle

\section{Introduction}

\IEEEPARstart{M}{otion} planning in automated driving has been extensively researched during the past years. Avoiding a collision in hazardous scenarios is particularly challenging on the operational and stability levels~\cite{Garrido2022}, i.e. planning a safe trajectory for the ego vehicle and tracking it during an emergency scenario. 

\begin{figure}[tb]
	\centering
	\includegraphics[width=0.48\textwidth]{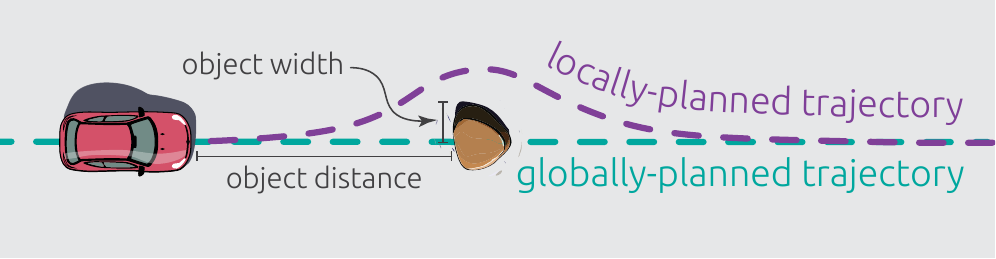}
	\caption{Elements of the collision avoidance problem after detecting a static obstacle}\label{fig:mooseprob}\vspace{-5mm}
\end{figure}

Among the various testing scenarios~\cite{Antkiewicz2020}, one example of an emergency situation is the sudden appearance of a static obstacle on the road, which can also be considered as the extreme case of leading vehicle deceleration in~\cite{DeGelder2021}. Figure~\ref{fig:mooseprob} shows a schematic view of influential elements for the planning problem in such scenarios: the distance to the object -- often reflected in time-to-collision or distance-to-collision threat measures in the literature~\cite{Dahl2019} -- and the width of the obstacle. For instance, shorter distance to the obstacle with a larger width can both contribute to the criticality of the situation as the width determines the necessary lateral displacement for collision avoidance. Understanding the importance of a wider width is closely linked to the current states of the vehicle as the required braking or steering actions for achieving a specific lateral displacement depend on factors like the current vehicle speed or sideslip angle.


MPC has become increasingly popular in the field of automated collision avoidance thanks to its straightforward handling of constraints and its capacity to dynamically adjust to environmental changes by solving the control optimization problem in a receding horizon manner~\cite{Stano2023}. In the current state of the art, trajectory planning and vehicle control are commonly addressed through one of two architectural frameworks: hierarchical~\cite{Fan2023,Yang2021-2,Chai2023,Qie2023,Wang2023,Wang2024,Dong2024} or integrated~\cite{Shang2023,Guo2018,Laurense2022,Li2023,Li2023-2,Nakka2023}. 

A hierarchical architecture offers greater flexibility in defining control problems and enables faster responses for real-time implementation, owing to the differing frequency and performance requirements at each level. However, the reference trajectory provided by the planner may not be attainable for the real plant. This is a critical issue in emergency cases where the feasible area for collision avoidance is limited and the vehicle is operating at its handling limits. Integrated planning and control circumvents this issue by treating both problems within a single optimization problem. 

While integrated MPC design, in particular when used with Electronic Power Steering (EPS), allows the handling of two optimization problems combined, it is essential to highlight a key distinction between the two architectures: in a hierarchical architecture, addressing the planning and control optimization problems can occur at different frequencies (e.g.~planning at 5-10Hz and control at 50-100Hz). Conversely, an integrated architecture demands solving the integrated optimization problem at the control frequency, albeit with a higher computational demand compared to the control optimization problem. 

An important source of computational complexity stems from the fact that the integrated MPC optimization problem is nonlinear, which requires employing computationally-demanding NonLinear Programming (NLP) algorithms to find the -- locally -- optimal solution. In~\cite{Li2023}, Gaussian safe envelopes used in the integrated MPC problem are obtained via Gaussian processes regression; this formulation allows for efficient solution of the integrated MPC problem by Quadratic Programming (QP). Other examples of QP solution of the MPC optimization problem include~\cite{Li2023-2} where constraining the decision space to the linear tire force range leads to a quadratic formulation of the problem, and~\cite{Sun2024} where the weights of the simplified quadratic problem are adapted online for improved control performance.

In~\cite{Laurense2022}, two models are serially cascaded to handle the two problems simultaneously, hence facilitating real-time solution of the optimization problem by NLP and a warm-start strategy. However, this strategy is limited in finding the optimal solution if a static obstacle suddenly appears on the road. Moreover, physics-based and local convexification~\cite{Wang2024,Sun2023} or explicit sub-optimal solution~\cite{Shang2023} have helped the real-time realization of integrated planning and control for normal driving. Yet, achieving real-time solutions in emergency scenarios remains a primary bottleneck of the current automated driving research. 

Another technique in this area is incorporating parametric approaches for trajectory planning into an integrated control optimization framework, which combines the strengths of both methods: while the integrated architecture prevents the generation of unattainable trajectories for the control layer, the parametric formulation streamlines optimization for faster convergence, which is essential for real-time planning and control. Parametric trajectories during a lane change maneuver are often modeled with polynomials at low speeds~\cite{Guo2018,Li2023-2,Wang2024,Dong2024} or sigmoid~\cite{Ammour2022} and tangent-hyperbolic functions~\cite{Sun2024} mimicking human behavior at high speeds~\cite{Yang2021}. For instance, when a static obstacle suddenly appears, the finite-state machine in~\cite{Ammour2022} triggers braking and steering for a lane change. The reference trajectory for such a maneuver is defined as a parametric sigmoid function and is real-time optimized via NLP. 

Despite the fast-paced progress of the literature, there remains a gap in achieving real-time implementation of automated collision-avoidance in real-life emergency situations. This is primarily due to the \emph{limited computational capacity of real systems}~\cite{Stano2023} and the low acceptance of the automated driving systems due to \emph{lack of interpretability}~\cite{Lu2024}. 

Given collision avoidance constraints, limited memory, and computation time, converging to -- even a local -- optimum via NLP is oftentimes not possible, which means that in practice, the best feasible solution found before hitting a stopping criteria is used. As a result, the `quality' of such a solution is sensitive to the provided initial guess. More specifically, the popular warm-start strategy which is often used in the literature, would be of limited value in such scenarios since the shifted solution of the previous time step after detecting the obstacle is often far from a feasible solution to avoid the detected obstacle. Limited computational resources hinder exploration of alternative starting points. As a result, having a ``good initial guess" is the practical way to improve the quality of the obtained solution by NLP. Here, a good initial guess is an \emph{interpretable} solution, mirroring a skilled driver's reaction. 

In this paper, we tackle the problem of real-time collision avoidance  after sudden appearance of a static obstacle with limited computational resources. The contributions of this paper are twofold:
\begin{enumerate}
    \item providing experimental insights into the real-time implementation of MPC-based collision avoidance by confronting the computational limitation challenge head-on in real-life scenarios using an electric vehicle, and
    \item improving the computational efficiency of the problem by integration of physics-based and human-like feed-forward planner, to help convergence to a feasible solution in emergencies. 
\end{enumerate}
We investigate multiple real-life emergency scenarios and analyze the effectiveness of our proposed approach in test cases that render previous solutions obsolete, emphasizing the need for our planning strategy. Hence, our approach integrates MPC more intelligently and provides effective solutions in the context of sudden appearance of a static obstacle.

The rest of this paper is structured as follows: Section~\ref{sec:vehicle} describes the vehicle model. followed by an overview of the proposed control system design in Section~\ref{sec:ctrlsys}. Section~\ref{sec:plan} then covers our proposed planning strategy. Details on system implementation are given in Section~\ref{sec:implement} and the results of the experimental tests are analyzed in Section~\ref{sec:results}. Finally, Section~\ref{sec:conclude} concludes this paper.

\section{Vehicle Model}\label{sec:vehicle}

\begin{figure}[H]
	\centering\vspace{-3mm}
	\includegraphics[width=0.35\textwidth]{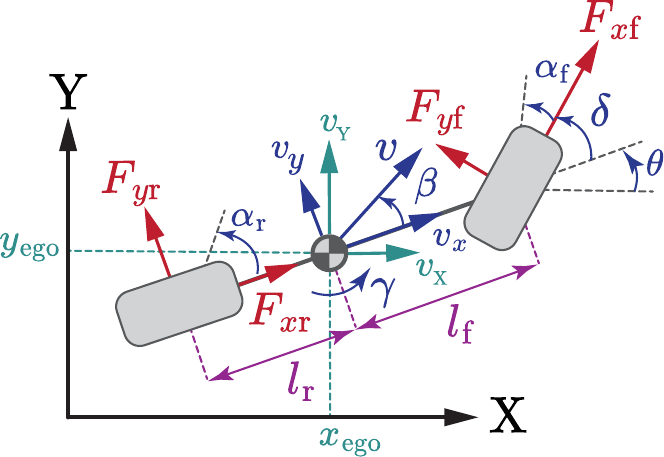}
	\caption{Single-track vehicle model}\label{fig:egomodel}\vspace{-3mm}
\end{figure}

The single track vehicle model is shown in Fig.~\ref{fig:egomodel}. First, we cover the kinematics of the vehicle and then we apply Newton's second law to derive the governing equations of motion. Table~\ref{tab:params} shows the variables and parameters used in this paper.

\subsection{Kinematic Model}

The trajectory of the ego vehicle in the global coordinates can be written as
\begin{subequations}\label{eq:trajdef}
\begin{align}
   \dot{x}\e &= v_X, \\
   \dot{y}\e &= v_Y, \\
   \dot{\theta} &= \gamma,
\end{align}
\end{subequations}
and the velocities in the local coordinate attached to the vehicle body are obtained by
\begin{subequations}\label{eq:trajdef}
\begin{align}
   v_x  &= v_X  \cos{\left(\theta \right)} - v_Y  \sin{\left(\theta \right)}, \\
   v_y  &= v_X  \sin{\left(\theta \right)} + v_Y  \cos{\left(\theta \right)}.
\end{align}
\end{subequations}
With the kinematic states and inputs defined by
\begin{align*}
    s\kin   =& \begin{bmatrix} x\e   & y\e  & v_X  & v_Y  \end{bmatrix}^T, \\
    u\kin   =& \begin{bmatrix} \dot{v}_X  & \dot{v}_Y \end{bmatrix}^T,
\end{align*}
the kinematic model of the vehicle can be written as
\begin{align}
    \dot{s}\kin = A s\kin + B u\kin ,\label{eq:skin}
\end{align}
where
\begin{align*}
    A = \begin{bmatrix} 0 & 0 & 1 & 0 \\ 0 & 0 & 0 & 1 \\ 0 & 0 & 0 & 0 \\ 0 & 0 & 0 & 0 \end{bmatrix}, &&
    B = \begin{bmatrix} 0 & 0 \\ 0 & 0 \\ 1 & 0 \\ 0 & 1 \end{bmatrix}.
\end{align*}
\begin{table}[h]
\centering
\caption{List of variables and parameters}
\label{tab:params}
\begin{center}
\begin{tabular}{c | l c }
    \noalign{\hrule height 0.3mm}
    Var.\ & Definition & Units \\
    \hline
    $x\e$       & Global longitudinal position of the CoG & m \\
    $y\e$       & Global lateral position of the CoG & m \\
    $x\obs$       & Global longitudinal position of the obstacle & m \\
    $y\obs$       & Global lateral position of the obstacle & m \\
    $v_X$       & Global longitudinal velocity of the CoG & m/s \\
    $v_Y$       & Global lateral velocity of the CoG & m/s \\
    $v_x$       & Local longitudinal velocity of the CoG & m/s \\
    $v_y$       & Local lateral velocity of the CoG & m/s \\
    $\theta$    & Yaw angle & rad \\
    $\beta$     & Sideslip angle & rad \\
    $\gamma$    & Yaw rate & rad/s \\
    $\delta$    & Steering angle & rad \\
    $\alf$      & Front slip angle & rad \\
    $\alr$      & Rear slip angle & rad \\
    $\Fxf$      & Longitudinal force of the front tire & N \\
    $\Fxr$      & Longitudinal force of the rear tire & N \\
    $\Fyf$      & Lateral force of the front tire & N \\
    $\Fyr$      & Lateral force of the rear tire & N \\
    $\Fzf$      & Normal force on the front tire & N \\
    $\Fzr$      & Normal force on the rear tire & N \\
    \noalign{\hrule height 0.3mm}
    Par.\ & Definition & Value \\
    \hline
    $\ts$       & Planning sampling time & 0.2s \\
    $m$         & Vehicle mass & 925kg \\
    $I_{zz}$    & Vehicle moment of inertia & 617kgm$^2$ \\
    $J_{\omega \mathrm{r}}$ & Rotational inertia of the rear wheel & 1.24kgm$^2$ \\
    $r$         & Wheel radius & 0.301m \\
    $\lf$       & Distance between the front axle and the CoG & 0.99m \\
    $\lr$       & Distance between the rear axle and the CoG & 0.71m \\
    $T_{\max}$  & Maximum torque & 200Nm \\
    $\delta_{\max}$ & Maximum steering angle & 0.3rad \\
    $\dot{\delta}_{\max}$ & Maximum steering angular speed & 2618rad/s \\
    $C_x$ & Longitudinal Pacejka tire constant & 1.5 \\
    $B_x$ & Longitudinal Pacejka tire & 8.0 \\
    $C_y$ & Lateral Pacejka tire constant & 1.4057 \\   
    $B_y$ & Lateral Pacejka tire & 7.1138 \\
    \noalign{\hrule height 0.3mm}   
\end{tabular}\vspace{-5mm}
\end{center}
\end{table}

\subsection{Equations of Motion}\label{subsec:eqmotion}
\begin{figure*}[htbp]
\begin{center}
\begin{tikzpicture}\sffamily\sansmath
	
    \node (mpc) [block] {\footnotesize MPC};

     \begin{pgfonlayer}{background}

        \path (mpc.west)+(-0.6,1.5) node (a) {};
        \path (a)+(+3.0*\edgedist,-4.2*\blockdist) node (b) {};
        \path [fill=TealBlue!20, draw=black,thick] (a) rectangle (b);
        
        \path (a)+(+1.5*\edgedist,+0.25*\blockdist) node {Planning};       

        \path (a)+(+3.1*\edgedist,0*\blockdist) node (f) {};
        \path (b)+(+2.6*\edgedist,0*\blockdist) node (g) {};  
        
        \path [fill=LimeGreen!30, draw=black,thick] (f) rectangle (g);
        \path (f)+(+1.25*\edgedist,+0.25*\blockdist) node {Motion Control};  

        \path (f)+(+2.6*\edgedist,-0.0*\blockdist) node (h) {};
        \path (g)+(+1.6*\edgedist,-0.0*\blockdist) node (i) {};  
        
        \path [fill=red!30, draw=black,thick] (h) rectangle (i);
        \path (h)+(+0.8*\edgedist,+0.25*\blockdist) node {System};
        
        \path (h)+(1.6*\edgedist,0.0*\blockdist) node (j) {};
        \path (i)+(1.6*\edgedist,0.0*\blockdist) node (k) {};  
        
        \path [fill=Lavender!30, draw=black,thick] (j) rectangle (k);
        \path (j)+(+0.6*\edgedist,+0.25*\blockdist) node {Perception};

    \end{pgfonlayer}

    \begin{pgfonlayer}{foreground}
        
        \path (mpc)+(0,-2*\blockdist) node (ff) [block] {\footnotesize MSF};
        \path (mpc)+(0.5*\edgedist,-1.0*\blockdist) node (weight) [block,text width=5.8em] {\footnotesize Weight function};
        \path (weight)+(1.5*\edgedist,-0*\blockdist) node (fblin) [block] {\footnotesize FBL};

        \path [draw, -Latex] (mpc) -| node [above,midway] {\scriptsize $[\dot{v}_X^\ast, \dot{v}^\ast_Y]\mpc$} ($(weight.north)+(0.2*\edgedist,0)$);
        \path [draw, -Latex] (ff) -| node [below,midway] {\scriptsize $[\dot{v}_X^\ast, \dot{v}^\ast_Y]\ff$} ($(weight.south)+(0.2*\edgedist,0)$);
        \draw [thick,-Latex] (weight.east) -- (fblin.west) node[ann,near end,below,xshift=0.5em] {\scriptsize $[\dot{v}_X^\ast, \dot{v}^\ast_Y]$};    
        \draw [densely dashed,-Latex] (weight)+(0.8*\edgedist,0) -- node [ann] {} + (0.8*\edgedist,1.8*\blockdist) node [ann] {} -| (mpc.north) node [above,near start] {\footnotesize initial guess};

        \path (f)+(0.4*\edgedist,-0.75*\blockdist) node (sumv) [sumprod] {};
        \node at ($(sumv.west)+(-0.1*\edgedist,+0.2*\blockdist)$) () {$+$};
        \path (sumv)+(0.6*\edgedist,-0.0*\blockdist) node (cvfb) [block] {\footnotesize $C_{v,\rm{fb}}$};
        \path (sumv)+(0.6*\edgedist,-1.0*\blockdist) node (cvff) [block] {\footnotesize $C_{v,\rm{ff}}$};
        \path (cvfb)+(0.5*\edgedist,-0.5*\blockdist) node (sumt) [sumprod] {};
        \node at ($(sumt.north)+(+0.1*\edgedist,+0.2*\blockdist)$) () {$+$};
        \node at ($(sumt.south)+(+0.1*\edgedist,-0.2*\blockdist)$) () {$+$};
        \path (sumt.east)+(0.5*\edgedist,-0*\blockdist) node (tdl) [block] {\footnotesize TDL};
        \path (tdl)+(0.0*\edgedist,-2.2*\blockdist) node (cdfb) [block] {\footnotesize $C_{\delta,\rm{fb}}$};
        \path (sumv)+(0.0*\edgedist,-2.7*\blockdist) node (sumd) [sumprod] {};
        \node at ($(sumd.west)+(-0.1*\edgedist,+0.2*\blockdist)$) () {$+$};
        \node at ($(sumd.south)+(-0.15*\edgedist,-0.2*\blockdist)$) () {$-$};

        \draw[thick,-Latex] (fblin.north) |- (sumv.west) node[ann,near start,right] {\scriptsize $v_x^\ast$};
        \draw[thick,-Latex] ($(sumv.west)+(-0.2*\edgedist,0)$) |- (cvff.west);
        \draw[thick,-Latex] (fblin.south) |- (sumd.west) node[ann,near start,right] {\scriptsize $\delta^\ast$};
        \draw[thick,-Latex] (sumv.east) -- (cvfb.west);
        \draw[thick,-Latex] (cvfb.east) -| (sumt.north);
        \draw[thick,-Latex] (cvff.east) -| (sumt.south);
        \draw[thick,-Latex] (sumt.east) -- (tdl.west);
        \draw[thick,-Latex] (sumd.east) -- (cdfb.west);
        
        \path (tdl)+(1.0*\edgedist,+0.55*\blockdist) node (iwmrl) [sumprod] {\scriptsize IWM${}_{\mathrm{rl}}$};
        \path (tdl)+(1.0*\edgedist,-0.55*\blockdist) node (iwmrr) [sumprod] {\scriptsize IWM${}_{\mathrm{rr}}$};
        \path (cdfb)+(1.0*\edgedist,-0.0*\blockdist) node (eps) [sumprod] {\footnotesize EPS};
        \path (iwmrl)+(0.7*\edgedist,-0.5*\blockdist) node (imu) [block,rotate=-90] {\footnotesize IMU};
        \path (imu)+(0.0*\edgedist,-2.0*\blockdist) node (gps) [block,rotate=-90] {\footnotesize GPS};

        \draw[thick,-Latex] (tdl.north) |- (iwmrl.west) node[above,midway] {\scriptsize $T_{\mathrm{rl}}$}; 
        \draw[thick,-Latex] (tdl.south) |- (iwmrr.west) node[below,midway] {\scriptsize $T_{\mathrm{rr}}$};
        \draw[thick,-Latex] (cdfb.east) -- (eps.west);
        \draw[thick,-Latex] (eps.south) |- ($(sumd.south)+(0,-0.4*\blockdist)$) -- (sumd.south) node[right,midway] {\scriptsize $\delta$};
        \draw[densely dashed] ($(fblin.east)+(0.3*\edgedist,-0.15*\blockdist)$) -- ($(fblin.east)+(2.6*\edgedist,-0.15*\blockdist)$) node[above,midway] {\scriptsize Velocity control} node[below,midway] {\scriptsize Steering angle control};

        \path (j)+(0.55*\edgedist,-1.3*\blockdist) node (observer) [block,text width=3.4em,minimum height=5.5em] {\footnotesize State observer};
        \path [fill=black,draw=black] ($(j)+(1.4*\edgedist,-0.2*\blockdist)$) rectangle ($(k)+(-0.2*\edgedist,+0.5*\blockdist)$);
        \path (observer)+(0.0*\edgedist,-2.2*\blockdist) node (detector) [block,text width=3.4em,minimum height=3em] {\footnotesize Obstacle detector};
        
        \draw[thick,-Latex] (imu.north) -- ($(observer.west)+(0,+0.1*\blockdist)$);
        \draw[thick,-Latex] (gps.west) |- ($(observer.west)+(0,-0.7*\blockdist)$);
        \draw[thick,-Latex] ($(eps.south)+(0,-0.1*\blockdist)$) -- ($(eps.south)+(0.4*\edgedist,-0.1*\blockdist)$) |- ($(observer.west)+(0.0*\edgedist,+0.85*\blockdist)$);

        \draw[thick,ultra thick,-Latex] ($(j)+(1.4*\edgedist,-2.2*\blockdist)$) -- ($(j)+(1.6*\edgedist,-2.2*\blockdist)$) |- ($(a)+(0.15,-4.4*\blockdist)$) |- (weight.west);   
        \draw[thick,ultra thick,-Latex] ($(a)+(0.15,-2.5*\blockdist)$) |- (mpc.west);
        \draw[thick,ultra thick,-Latex] ($(a)+(0.15,-2.5*\blockdist)$) |- (ff.west);

        \draw[thick,-Latex] ($(observer.east)+(0.0,+0.9*\blockdist)$) -- ($(observer.east)+(0.4*\edgedist,+0.9*\blockdist)$) node[above,midway] {\scriptsize $\hat{v}_x$};
        \draw[thick,-Latex] ($(observer.east)+(0.0,+0.45*\blockdist)$) -- ($(observer.east)+(0.4*\edgedist,+0.45*\blockdist)$) node[above,midway] {\scriptsize $\hat{v}_y$};
        \draw[thick,-Latex] ($(observer.east)+(0.0,+0.0*\blockdist)$) -- ($(observer.east)+(0.4*\edgedist,+0.0*\blockdist)$) node[above,midway] {\scriptsize $\hat{\beta}$};
        \draw[thick,-Latex] ($(observer.east)+(0.0,-0.45*\blockdist)$) -- ($(observer.east)+(0.4*\edgedist,-0.45*\blockdist)$) node[above,midway] {\scriptsize $\hat{\gamma}$};
        \draw[thick,-Latex] ($(observer.east)+(0.0,-0.9*\blockdist)$) -- ($(observer.east)+(0.4*\edgedist,-0.9*\blockdist)$) node[above,midway] {\scriptsize $\hat{\theta}$};

        \draw[thick,-Latex] (observer.south) -- (detector.north) node[right,midway] {\scriptsize $[x\e, y\e]$};
        \draw[thick,-Latex] ($(detector.east)+(0.0,+0.0*\blockdist)$) -- ($(detector.east)+(0.4*\edgedist,+0.0*\blockdist)$) node[above,near start,xshift=0.4em] {\scriptsize $[\tau, \nu]$};

        \draw[thick,-Latex] ($(observer.east)+(0.1*\edgedist,+0.9*\blockdist)$) -- ($(observer.east)+(0.1*\edgedist,+0.9*\blockdist)+(0.0*\edgedist,+0.25*\blockdist)$) -| (sumv.north) node[left,near end] {\scriptsize $-$};
        
        
    \end{pgfonlayer}
		
\end{tikzpicture}
\end{center}
\caption{The proposed control architecture}\label{fig:diagraml}
\end{figure*}
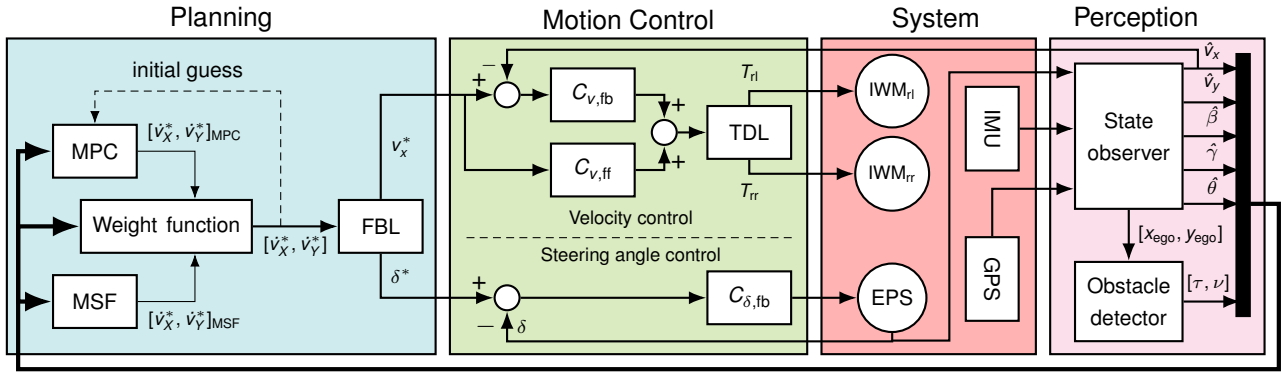 
The governing equations of motion for the vehicle model in Fig.~\ref{fig:egomodel} can be written as follows~\cite{Chowdhri2021}:
\begin{subequations}\label{eq:eom}
\begin{align}
   \sum F_x &= \Fxf \cos{(\delta)} - \Fyf \sin{(\delta)} + \Fxr,  \\ 
   \sum F_y &= \Fxf \sin{(\delta)} + \Fyf \cos{(\delta)} + \Fyr, \\
   \sum M_z &= \left(\Fxf \sin{(\delta)} + \Fyf \cos{(\delta)}\right) \lf - \Fyr \lr.
\end{align}
\end{subequations}
Considering a small front steering angle, the dynamics of the single-track vehicle model is obtained as
\begin{subequations}\label{eq:eomdyn}
\begin{align}
   \dot{v}_x =& \; \dfrac{1}{m} \Big(\Fxf + \Fxr \Big) + v_y \gamma, \label{eq:dvxnl} \\
   \dot{v}_y =& \; \dfrac{1}{m} \Big( \Fyf + \Fyr \Big) - v_x \gamma, \label{eq:dvynl} \\
   \dot{\gamma} =& \; \dfrac{1}{I_{zz}} \Big( \Fyf  \; \lf - \Fyr \;\lr \Big).\label{eq:dgammanl}
\end{align}
\end{subequations}
Since we expect that the evasive maneuver will generate forces beyond their maximum peak~\cite{Galvani2014}, we consider a nonlinear model for the tire forces on the front and rear axles using the celebrated Pacejka tire model~\cite{Pacejka2005} as
\begin{subequations}\label{eq:fy}
	\begin{align}
		\Fyf &= \mu \Fzf \sin \left( C_y \arctan \left( B_y \alf \right)\right),\label{eq:fyf}\\
		\Fyr &= \mu \Fzr \sin \left( C_y \arctan \left( B_y \alr \right)\right),\label{eq:fyr}
	\end{align}
\end{subequations}
with the front and rear slip angles respectively described by
\begin{subequations}\label{eq:alphas}
	\begin{align}
		&\alf = \delta - \beta - \dfrac{\lf \gamma}{v_x}, \label{eq:alf}\\
		&\alr = \dfrac{\lr \gamma}{v_x} - \beta. \label{eq:alr}
	\end{align}
\end{subequations}

\section{Proposed Control System}\label{sec:ctrlsys}

Figure~\ref{fig:diagraml} shows the architecture of the proposed closed-loop system, consisting of the real system, as well as perception, planner and controller modules. 

\paragraph{Perception} On the perception layer, an observer is incorporated to estimate the sideslip and yaw angles, while the obstacle detector obtains the position and width of the obstacle and returns the time to collision $\tau$, and lateral steering index $\nu$, based on the measurements of the current states. The lateral steering index indicates the degree of extreme steering required to avoid a collision, and is covered in the next section in more detail.

\paragraph{Planner} We propose a planner design strategy that leverages the capabilities of MPC in finding an optimal maneuvers for collision avoidance, as well as an assistive Maximum Steering Feed-forward (MSF) planner designed to replicate a human-like response to the detection of static obstacles denoted by maximum steering maneuver. For the sake of computational efficiency, the kinematic model of the ego vehicle is used in the planner module. Therefore, both MPC and MSF planners provide their respective references for the longitudinal and lateral accelerations, $\dot{v}_x^\ast$ and $\dot{v}_y^\ast$. A weight function then combines the two references and feeds the resulting reference signals to a FeedBack Linearization (FBL) controller to obtain the corresponding reference steering angle and longitudinal velocity.

\paragraph{Controller} We use a SPeed Controller (SPC) to obtain the required torque for tracking the reference longitudinal velocity. The output torque of the speed controller is then distributed via the Torque Distribution Law (TDL) to the rear left and rear right In-Wheel Motors (IWMs). In addition, the FeedBacked Electric Power Steering (FB-EPS) system tracks the steering reference by providing the steering angle signal to the steering motor.

\section{Design of the Planning Strategy}\label{sec:plan}

Given the measured and observed current states of the ego vehicle, along with $\tau$ and $\nu$, the MPC and MSF planners provide their respective solutions, $\dot{v}_X^\ast$ and $\dot{v}_Y^\ast$, distinguished by their respective subscripts in Fig.~\ref{fig:diagraml}. The same information is fed into the weight function that acts as a situation-assessment unit by distributing the weights between $[\dot{v}_X^\ast,\dot{v}_Y^\ast]\mpc$ and $[\dot{v}_X^\ast,\dot{v}_Y^\ast]\ff$. For instance, in cases where the solution of the MPC planner is close to the steering limits or if it fails to provide a solution in time, the weight function assigns a higher weight to solution from the MSF planner. The design of each planner is discussed separately in the following sections.

\subsection{MSF Planner}

We define the maximum steering maneuver \mbox{$y_{\max} : \mathbb{R}^2 \to \mathbb{R}$} to denote the resulting lateral position of the vehicle, as a function of time and its longitudinal velocity, after performing a maximum-possible evasive action without braking. This evasive steering is designed to mimic one of the human-like responses after detecting an obstacle: steering to the side as fast as possible. 
\begin{figure}[hbtp]
\begin{center}\vspace{-2mm}
\begin{tikzpicture}[declare function={yego(\v,\y) =-0.8001 + 0.4752*\v + 2.1603*\y -0.0481*\v^2 + -1.1421*\v*\y + -1.3265*\y^2 + 0.0846*\v^2*\y +  0.6708*\v*\y^2 +  0.2326*\y^3 -0.0183*\v^2*\y^2 + -0.0592*\v*\y^3 + -0.0134*\y^4;},]
	\begin{axis}[height=0.25\textwidth,width=0.5\textwidth,
		xlabel=$t$ (s),ylabel=$y_{\max}$ (m),xmin=0,xmax=5,ymin=0,ymax=25,legend pos=north west,
		legend style={draw=none, fill opacity=0.0, text opacity = 1}]	
		\addplot[color=BrickRed,smooth,thick] table [x=Time,y=VX01]{Data/Avoidance.dat};	
		\addplot[color=BurntOrange,smooth,thick] table [x=Time,y=VX03]{Data/Avoidance.dat};	
		\addplot[color=TealBlue,smooth,thick] table [x=Time,y=VX05]{Data/Avoidance.dat};	
		\addplot[color=MidnightBlue,smooth,thick] table [x=Time,y=VX07]{Data/Avoidance.dat};	
		\legend{$v_x = 1$ m/s,$v_x = 3$ m/s,$v_x = 5$ m/s,$v_x = 7$ m/s};
	\end{axis}
\end{tikzpicture}
\caption{Maximum steering maneuver (\ref{eq:ymax}) for different longitudinal velocities.}\label{fig:path}\vspace{-2mm}
\end{center}
\end{figure}
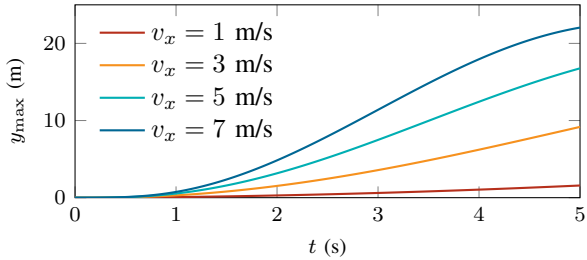

The function $y_{\max}$ is calculated for time $t$ and velocity $v_x$ by considering (\ref{eq:trajdef})-(\ref{eq:alphas}) to solve
\begin{subequations}\label{eq:ymax}
\begin{align}
    y_{\max} (t,v_x)  & = \int_0^t v_Y (\phi) d\phi,\\
    \mathrm{s.t.} \quad   v_x (\phi) & = v_x, \\
    \delta (\phi) & = \min \left( \dot{\delta}_{\max}  \phi\; , \; {\delta}_{\max} \right),\\
    v_y (0) & = 0,\\
    \theta (0) & = 0,\\
    \gamma (0) & = 0,
\end{align}
\end{subequations}
where $v_Y$ is obtained by solving $\dot{\delta}_{\max}$ and ${\delta}_{\max}$ respectively represent the maximum steering rate and steering angles. The resulting lateral position is shown by solid lines in Fig.~\ref{fig:path}. Using $y_{\max}$ and the width of the danger zone $w$, the lateral steering index $\nu$ is defined by
\begin{align}\label{eq:nudef}
	\nu = \dfrac{w}{y_{\max} (\tau, v_x)}.
\end{align}
\begin{rem}
    The zero initial conditions for the lateral velocity, yaw angle, and yaw rate in (\ref{eq:ymax}) represent a specific and extreme case. To define a more realistic model of the maximum steering maneuver, $y_{\max}$ can be defined as a function of these initial conditions. However, this will lead to a higher-dimensional domain for the function $y_{\max}$, hence resulting in higher computational demand in the next steps. For the sake of computational efficiency, we use the zero initial conditions to merely account for the most extreme form of the maximum steering maneuver.
\end{rem}
\subsection{MPC Planner}

We define the safety barrier $y\barr$ using the Sigmoid barrier function in~\cite{Ammour2022} in its extreme case as
\begin{align}\label{eq:ysafe}
	y\barr (k) = \dfrac{w}{1 + \exp \left(\dfrac{x\obs - x\e (k) - 4 v_X (k) }{\epsilon \sqrt{w \; v^2_X (k)}} \right)},
\end{align}
where $w$ is the width of the danger zone, to be avoided by the ego vehicle's center of gravity and \mbox{$\epsilon = 1/ \sqrt{8.8 \mu g}$}. In the following example, we clarify the behavior of the $y\barr$ function. 
\begin{exmp}
    Figure~\ref{fig:barrier} shows a schematic view of the Sigmoid barrier function (\ref{eq:ysafe}) with the danger zone shown in red. For a given $v_X$, $y\barr$ is defined such that the vehicle's center of gravity has traveled $w/2$ in the lateral direction 4 seconds before arriving at the longitudinal position of the danger zone. For instance, assuming a constant longitudinal speed for the vehicle in the global coordinates, with $\tau=6$s shown in blue, $y\barr$ passes through the lateral mid-point after 2 seconds. If the obstacle is closer, i.e.~$\tau=5$s as shown in orange, the resulting $y\barr$ has the same curvature and ensures a $4 v_x$ distance before reaching $w/2$.    
    \begin{figure}[t]\centering
    \begin{tikzpicture}
        \begin{axis}[height=0.25\textwidth,width=0.3\textwidth,ytick=\empty,xtick=\empty,clip=false,ymajorgrids,
        xlabel=$X$,ylabel=$Y$,xmin=0,xmax=8,ymin=-1,ymax=2.6,legend pos=south east,axis x line=middle,axis y line=middle,legend style={draw=none, fill opacity=0.0, text opacity = 1}]       
            \draw[red,fill=red] (6,0) -- (6,2) -- (7,2) -- (7,0) -- (6,0);
            \addplot[color=black, dashed] coordinates {(0,2) (5,2)};
            \node[left] at (0.0,2) {$w$};
            \addplot[color=black, dashed] coordinates {(2,1) (2,0)};
            \draw[black,Latex-Latex] (0,1) -- (2,1) node[above,midway] {$2 v_x$};
            \draw[black,Latex-Latex] (2,1) -- (6,1) node[above,midway] {$4 v_x$};
            \addplot[color=NavyBlue,smooth,thick,samples=50,domain=0:7]{2/(1+exp(-(4*1+x-6)*6.6/(1))}; 
        \end{axis}
    \end{tikzpicture}
    \begin{tikzpicture}
        \begin{axis}[height=0.25\textwidth,width=0.3\textwidth,ytick=\empty,xtick=\empty,clip=false,ymajorgrids,
        xlabel=$X$,ylabel=$Y$,xmin=0,xmax=8,ymin=-1,ymax=2.6,legend pos=south east,axis x line=middle,axis y line=middle,legend style={draw=none, fill opacity=0.0, text opacity = 1}]       
            \draw[red,fill=red] (5,0) -- (5,2) -- (6,2) -- (6,0) -- (5,0);
            \addplot[color=black, dashed] coordinates {(0,2) (4,2)};
            \node[left] at (0.0,2) {$w$};
            \addplot[color=black, dashed] coordinates {(1,1) (1,0)};
            \draw[black,Latex-Latex] (0,1) -- (1,1) node[above,midway] {$v_x$};
            \draw[black,Latex-Latex] (1,1) -- (5,1) node[above,midway] {$4 v_x$};
            \addplot[color=Apricot,smooth,thick,samples=50,domain=0:7]{2/(1+exp(-(4*1+x-5)*6.6/(1))}; 
        \end{axis}
    \end{tikzpicture}
    \caption{Sigmoid barrier function examples}\label{fig:barrier}\vspace{-2mm}
    \end{figure}
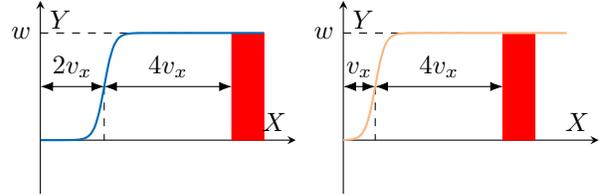
\end{exmp}

Considering a prediction horizon of length $\Np$, we denote the input signal and the resulting predicted states over the prediction window by 
\begin{subequations}
\begin{align}
\tilde{s}\kin (k) &= \begin{bmatrix} s\kin^T (k+1 | k) & \dots & s\kin^T (k+\Np | k)\end{bmatrix}^T,\label{eq:stilde}\\
	\tilde{u}\kin (k) &= \begin{bmatrix} u\kin^T (k) & \dots & u\kin^T (k+\Np-1)\end{bmatrix}^T,\label{eq:utilde}
\end{align} 
\end{subequations}
where $s\kin (k+i | k)$ for \mbox{$i \in \{1, \dots, \Np\}$} represents the predicted state at time step $k+i$ based on the state measurement at time step $k$, i.e.~$s\kin(k|k)$. We define four costs for each predicted step as
\begin{subequations}\label{eq:costs}
	\begin{align}
		J\barr (k) =& \; - \log \left( y\barr (k) -  y\e (k) \right), \label{eq:jbarr}\\
		J\stab (k) =& \; \left\vert v_Y (k) \left/ v_X (k) \right. \right\vert, \label{eq:jstab}\\
		J\vel (k) =& \;   \left\vert v_X (k) - v\des (k) \right\vert,\label{eq:jvel}\\
		J\steer (k) =& \; \left\vert \dot{v}_Y (k) \right\vert. \label{eq:jsteer}
	\end{align}
\end{subequations}
The function $J\barr$ represents the barrier function for safe collision avoidance and the functions $J\stab$, $J\vel$, and $J\steer$ are defined to respectively minimize the sideslip angle for vehicle stability, deviation from the desired velocity $v\des$, and steering action in the planning optimization problem. The MPC cost function is then defined as
\begin{equation}\label{eq:jcaldef}
	\begin{split}
		\Jcal (k) = \sum_{i = 1}^{\Np} \; \Big[ \; \eta_1 \; J\barr (k+i) + \eta_2 \; J\stab (k+i) \\
		+ \eta_3 \; J\vel (k+i) + \eta_4 \; J\steer(k+i) \Big],
	\end{split}
\end{equation}
where the weights \mbox{$0 \leqslant \eta_j \leqslant 1$} with $j \in \{1, \dots 4\}$ is a convex combination where \mbox{$\sum_1^4 \eta_j = 1$} and should be selected based on priorities in the test scenario; we elaborate more on this in Example~\ref{emp:eta}. The resulting MPC planning optimization problem is given by
\begin{align}
    \tilde{u}^\ast\kin (k) = \argmin\limits_{\tilde{u}\kin (k)} \quad \Jcal (k).  \label{eq:planmpc}%
\end{align}
Note that the decision variable in (\ref{eq:planmpc}) is the input vector and the state trajectory is defined within the cost function to get an unconstrained MPC optimization problem and avoid infeasibility issues. MPC finds the optimal trajectory to avoid a collision by solving (\ref{eq:planmpc}) in each time step in a receding-horizon fashion. This is done by solving the problem for the next $\Np$ time steps, while providing the solution to the first step ahead to the controller. 
\begin{exmp}\label{emp:eta}
    Consider a simple case of $\Np=2$ with $u\kin (k) = u\kin (k+1)$, \mbox{$v_X (k) = v\des (k) = 5$m/s}, \mbox{$x\e (k) = y\e (k) = y\obs (k) = 0$m}, \mbox{$x\obs = w = 1 $m} and \mbox{$v_Y (k) = 0$m/s}. The solution to (\ref{eq:planmpc}) for input signals normalized on the bound $[-1,1]$ with the selections of $\eta_j = 0.25$ is obtained as follows:
    \begin{align*}
        \begin{rcases}
            \eta_1 = 0.25, & \eta_2 = 0.25 \\
            \eta_3 = 0.25, & \eta_4 = 0.25
        \end{rcases}        
        &\implies 
        u\kin^\ast (k) = \begin{bmatrix}
            0.00 \\ 0.53
        \end{bmatrix},
    \end{align*}
    which indicates the optimal response would be no braking and steering rate equal to 53\% of its maximum value. However, if the weight for $J\barr$ is increased at the expense of reduction of the weights for $J\vel$ and $J\steer$, the solution to (\ref{eq:planmpc}) would change to
    \begin{align*}
        \begin{rcases}
            \eta_1 = 0.50, & \eta_2 = 0.25 \\
            \eta_3 = 0.05, & \eta_4 = 0.20
        \end{rcases}        
        &\implies 
        u\kin^\ast (k) = \begin{bmatrix}
            -0.35 \\ 1.00
        \end{bmatrix},
    \end{align*}
    corresponding to maximum steering rate with 35\% of maximum braking.
\end{exmp}

\subsection{FBL Controller}

To extract the reference steering angle and longitudinal velocity from \mbox{$[\dot{v}_X^\ast, \dot{v}^\ast_Y]$}, we obtain the required velocities in the local coordinate using (\ref{eq:trajdef}) and consider the vehicle dynamics in a feedback-linearization fashion as 
\begin{subequations}
	\begin{align}
			{\delta}^\ast (k) \; = \; & \beta (k) + \lf \gamma (k) \left/ \right. v_x (k), \notag \\
			& + \Fyf^{-1} \Big[ m \Big(\dot{v}_y (k)  + \gamma (k) v_x (k)\Big) \\
			& - \Fyr^{-1} \Big(\lr \gamma (k) \left/ \right. v_x (k) - \beta(k)\Big)  \Big] \notag \\
			{v_x}^\ast (k) \; = \; & \dot{v}_x (k) \ts + v_x (k).
	\end{align}
\end{subequations}
\begin{rem}
    In order to perform the proposed control algorithm, it is essential to know the tire model parameters and vehicle body sideslip angle. Using novel lateral tire force sensors available at our research group, the lateral tire force model can be identified~\cite{Nam2012}. Utilizing the merit of in-wheel-motor actuators, the road friction coefficient can be obtained without special difficulty~\cite{Furukawa2003}. Furthermore, we developed several methods to estimate the sideslip angle and yaw angle almost accurately in real-time~\cite{Nam2012,Wang2014,Nguyen2013}. Such estimation methods are not presented, as they are not the main goal of this paper.
\end{rem}

\section{System Implementation}\label{sec:implement}

\begin{figure}[hbtp]
	\centering
	\includegraphics[width=0.48\textwidth]{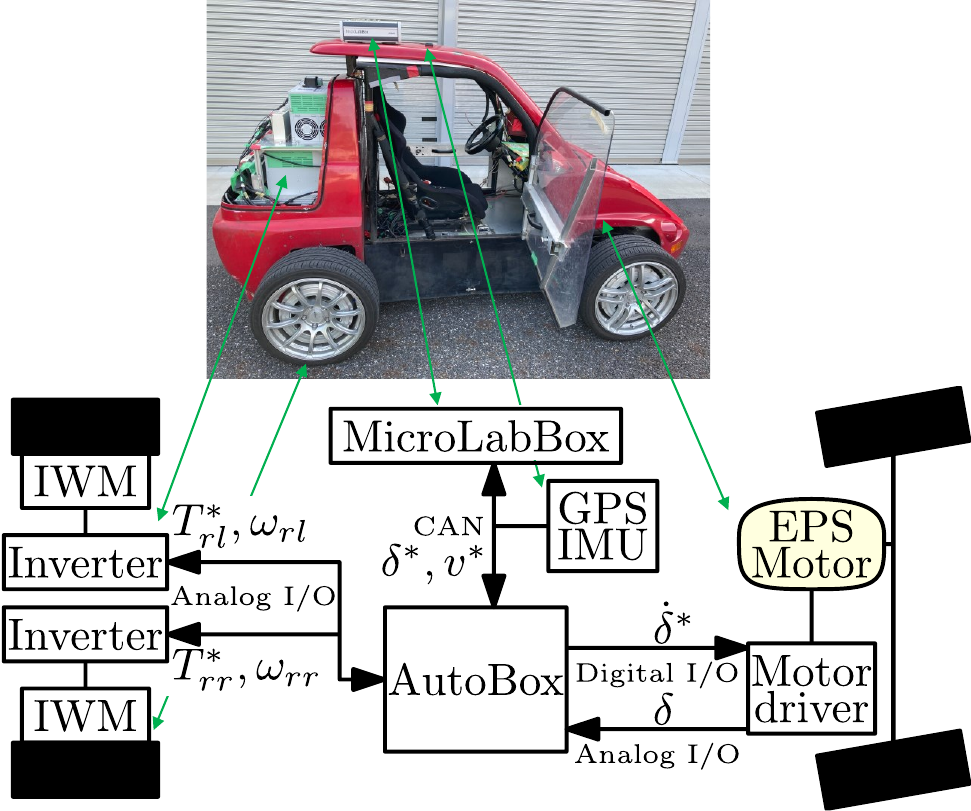}
	\caption{Experimental vehicle system configuration.}\label{fig:expvehiclesystem}
\end{figure}

To evaluate the proposed planning strategy, we use an experimental vehicle FPEV2-Kanon driven by rear-IWMs shown in Fig.~\ref{fig:expvehiclesystem}. The main parameters of the vehicle and IWM are summarized in Table \ref{tab:params}. FPEV2-Kanon is equipped with an electric power steering system consisting of a speed-controlled Maxon brushless motor. 

The motion control and planning modules are implemented in 
AutoBox (PPC 750GX 1 GHz, 32GB SDRAM program memory, 96MB SDRAM data storage) and MicroLabBox (NXP QorlQ P5020 dual core 2GHz, 1GB DRAM, 128MB flash memory), respectively. The communication among these modules and on-board Inertial Measuring Unit (IMU) and Global Positioning System (GPS) sensors is facilitated via Controller Area Network (CAN), and set to 500kbps. GPS, Autobox, and  MicroLabBox measurements are respectively updated at 1 Hz, 10kHz, and 5Hz frequencies. The MPC planner is implemented using \texttt{fmincon}'s SQP solver from \textsc{Matlab} R2017b Optimization toolbox. The optimization parameters are shown in Table \ref{table:op_para}.

The motion control module consists of two separate controllers for tracking the steering angle and the longitudinal velocity, as shown in Fig.~\ref{fig:diagraml}. Considering the nominal plant transfer function 
\begin{align*}
    P_{\rm{n}} (s) =  \dfrac{1}{\Big(J_{\omega \mathrm{r}}+r^{2}\frac{m}{2}(1-\lambda_{\rm{n}})\Big) s},
\end{align*}
with the nominal slip ratio \mbox{$\lambda_{\rm{n}} = 0.05$}, the Proportional Integral (PI) speed controller is designed by the placing the pole in $-1$rad/s.

\begin{table}[tb]
    \centering
    \caption{Optimization implementation parameters.}
    \label{table:op_para}
    \begin{center}
      \begin{tabular}{ccc}
        \noalign{\hrule height 0.3mm}
        Parameter & Value \\
        \hline
        Max iteration & 10 \\
        Max function evaluation & 100 \\
        Constraint tolerance & 1e-3 \\
        Optimality tolerance & 1e-3 \\
        Step tolerance & 1e-3 \\
        \noalign{\hrule height 0.3mm}
      \end{tabular}
    \end{center}\vspace{-2mm}
\end{table}

\begin{rem}
    Maximum number of iterations and function evaluations are set so that the program execution in MicroLabBox would not stop due to memory and computational burden.
\end{rem}

\section{Results and Discussion}\label{sec:results}

In this section, the results are discussed in three categories: the relaxed cases with $v_x = 5$m/s, lateral assessment with $v_x = 6$m/s, and extreme cases where $v_x = 7$m/s.

\subsection{Relaxed cases with $v_x = 5$m/s}

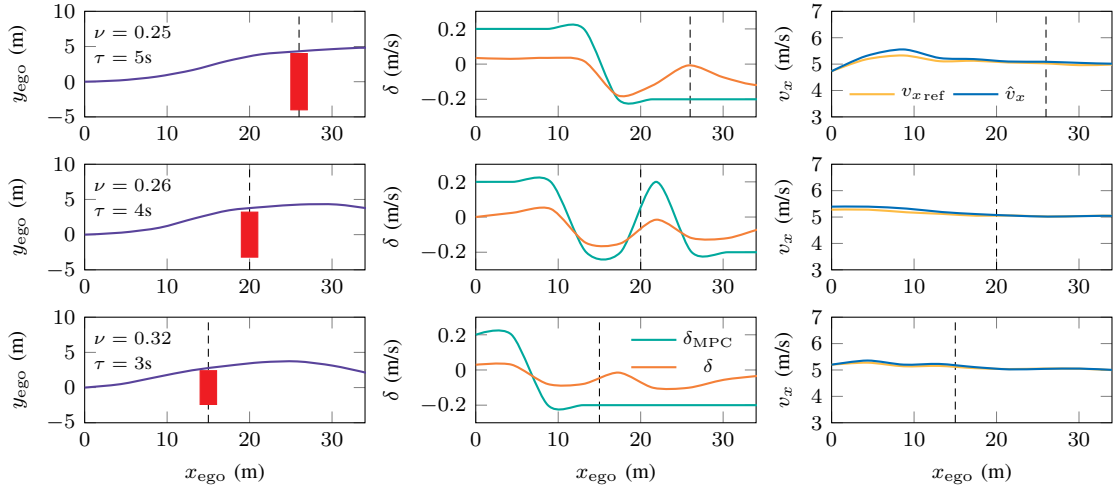
\begin{figure*}[hbtp]\centering
\begin{tikzpicture}
	\begin{axis}[height=0.18\textwidth,width=0.32\textwidth,
		ylabel=$y\e$ (m),xmin=0,xmax=34,ymin=-5,ymax=10]	
		\draw[densely dashed] (26,-5)--(26,10);
		\draw[Red,fill=Red,] (25,-4) rectangle (27,4);
		\addplot[color=Violet,smooth,thick] table [x=X17,y=Y17]{Data/Experiments.dat};
		\node[right] at (0,7) {\scriptsize $\nu = 0.25$};
		\node[right] at (0,3.5) {\scriptsize $\tau = 5$s};
	\end{axis}
\end{tikzpicture}
\begin{tikzpicture}
	\begin{axis}[height=0.18\textwidth,width=0.32\textwidth,
		ylabel=$\delta$ (m/s),xmin=0,xmax=34,ymin=-0.3,ymax=0.3]	
		\draw[densely dashed] (26,-0.3)--(26,0.3);
		\addplot[color=Emerald,smooth,thick] table [x=X17,y=UMPC17]{Data/DetailedCases.dat};
		\addplot[color=Orange,smooth,thick] table [x=X17,y=UOUT17]{Data/DetailedCases.dat};
	\end{axis}
\end{tikzpicture}
\begin{tikzpicture}
	\begin{axis}[height=0.18\textwidth,width=0.32\textwidth,
        legend style={draw=none, fill opacity=0.0, text opacity = 1},legend columns=2, legend pos=south west,
		ylabel=$v_x$ (m/s),xmin=0,xmax=34,ymin=3,ymax=7]	
		\draw[densely dashed] (26,3)--(26,7);
		\addplot[color=Dandelion,smooth,thick] table [x=X17,y=VREF17]{Data/DetailedCases.dat};
		\addplot[color=NavyBlue,smooth,thick] table [x=X17,y=VREAL17]{Data/DetailedCases.dat};
        \legend{\scriptsize ${v_x}\rf$,\scriptsize $\hat{v}_x$};
	\end{axis}
\end{tikzpicture}
\begin{tikzpicture}
	\begin{axis}[height=0.18\textwidth,width=0.32\textwidth,
		ylabel=$y\e$ (m),xmin=0,xmax=34,ymin=-5,ymax=10]	
		\draw[densely dashed] (20,-5)--(20,10);
		\draw[Red,fill=Red,] (19,-3.2) rectangle (21,3.2);
		\addplot[color=Violet,smooth,thick] table [x=X16,y=Y16]{Data/Experiments.dat};
		\node[right] at (0,7) {\scriptsize $\nu = 0.26$};
		\node[right] at (0,3.5) {\scriptsize $\tau = 4$s};
	\end{axis}
\end{tikzpicture}
\begin{tikzpicture}
	\begin{axis}[height=0.18\textwidth,width=0.32\textwidth,
		ylabel=$\delta$ (m/s),xmin=0,xmax=34,ymin=-0.3,ymax=0.3]	
		\draw[densely dashed] (20,-0.3)--(20,0.3);
		\addplot[color=Emerald,smooth,thick] table [x=X16,y=UMPC16]{Data/DetailedCases.dat};
		\addplot[color=Orange,smooth,thick] table [x=X16,y=UOUT16]{Data/DetailedCases.dat};
	\end{axis}
\end{tikzpicture}
\begin{tikzpicture}
	\begin{axis}[height=0.18\textwidth,width=0.32\textwidth,
		ylabel=$v_x$ (m/s),xmin=0,xmax=34,ymin=3,ymax=7]	
		\draw[densely dashed] (20,3)--(20,7);
		\addplot[color=Dandelion,smooth,thick] table [x=X16,y=VREF16]{Data/DetailedCases.dat};
		\addplot[color=NavyBlue,smooth,thick] table [x=X16,y=VREAL16]{Data/DetailedCases.dat};
	\end{axis}
\end{tikzpicture}
\begin{tikzpicture}
	\begin{axis}[height=0.18\textwidth,width=0.32\textwidth,
		xlabel=$x\e$ (m),ylabel=$y\e$ (m),xmin=0,xmax=34,ymin=-5,ymax=10]	
		\draw[densely dashed] (15,-5)--(15,10);
		\draw[Red,fill=Red,] (14,-2.4) rectangle (16,2.4);
		\addplot[color=Violet,smooth,thick] table [x=X02,y=Y02]{Data/Experiments.dat};
		\node[right] at (0,7) {\scriptsize $\nu = 0.32$};
		\node[right] at (0,3.5) {\scriptsize $\tau = 3$s};
	\end{axis}
\end{tikzpicture}
\begin{tikzpicture}
	\begin{axis}[height=0.18\textwidth,width=0.32\textwidth,
        legend style={draw=none, fill opacity=0.0, text opacity = 1,},legend columns=1, legend pos=north east,
		xlabel=$x\e$ (m),ylabel=$\delta$ (m/s),xmin=0,xmax=34,ymin=-0.3,ymax=0.3]	
		\draw[densely dashed] (15,-0.3)--(15,0.3);
		\addplot[color=Emerald,smooth,thick] table [x=X02,y=UMPC02]{Data/DetailedCases.dat};
		\addplot[color=Orange,smooth,thick] table [x=X02,y=UOUT02]{Data/DetailedCases.dat};
        \legend{\scriptsize $\delta\mpc$, \scriptsize $\delta$};
	\end{axis}
\end{tikzpicture}
\begin{tikzpicture}
	\begin{axis}[height=0.18\textwidth,width=0.32\textwidth,
		xlabel=$x\e$ (m),ylabel=$v_x$ (m/s),xmin=0,xmax=34,ymin=3,ymax=7]	
		\draw[densely dashed] (15,3)--(15,7);
		\addplot[color=Dandelion,smooth,thick] table [x=X02,y=VREF02]{Data/DetailedCases.dat};
		\addplot[color=NavyBlue,smooth,thick] table [x=X02,y=VREAL02]{Data/DetailedCases.dat};
	\end{axis}
\end{tikzpicture}
\caption{Experimental assessment of the $\tau$ influence: vehicle trajectory for the desired velocity $v_x = 5$m/s.}\label{fig:exptx5}
\end{figure*}

Figure~\ref{fig:exptx5} shows the results for the first round of experimental tests. The vehicle accelerates to reach the longitudinal velocity of $5$m/s and then detects a static obstacle with \mbox{$\tau \in \{3,4,5\}$} seconds. The desired vehicle speed is set to $5$m/s as well, and to compensate for the lower $\tau$ values, we select a smaller obstacle for lower $\tau$ values to keep the $\nu$ parameter in a narrow band. In all three scenarios, the vehicle manages to safely avoid colliding with the static obstacle with the help of the MPC and feed-forward steering commands while keeping the velocity close to the desired value. In all three cases, the maximum steering command provided by the combined planner occurs after passing the obstacle. At this point, the planner decides to steer back to the initial lateral position since there is no longer a risk of colliding with the obstacle. However, the communication delays in reading the GPS signal result in a slight delay in responding as well, as can be observed in the cases with $\tau=4$s and $3$s.
\begin{obsrv} 
    While MPC is prone to providing solutions close to the steering bounds in limited computation time, the combined planning strategy offers a smoother maneuver.
\end{obsrv}

\subsection{Lateral assessment with $v_x = 6$m/s}

\begin{figure*}[hbtp]\centering
\begin{tikzpicture}
	\begin{axis}[height=0.18\textwidth,width=0.32\textwidth,
		ylabel=$y\e$ (m),xmin=0,xmax=34,ymin=-5,ymax=10]	
		\draw[densely dashed] (20,-5)--(20,10);
		\draw[Red,fill=Red,] (19,-2) rectangle (21,2);
		\addplot[color=Violet,smooth,thick,] table [x=X18,y=Y18]{Data/Experiments.dat};
		\node[right] at (0,7) {\scriptsize $\nu = 0.21$};
		\node[right] at (0,3.5) {\scriptsize $\tau = 3$s};
	\end{axis}
\end{tikzpicture}
\begin{tikzpicture}
	\begin{axis}[height=0.18\textwidth,width=0.32\textwidth,
		ylabel=$\delta$ (m/s),xmin=0,xmax=34,ymin=-0.3,ymax=0.3]	
		\draw[densely dashed] (20,-0.3)--(20,0.3);
		\addplot[color=Emerald,smooth,thick] table [x=X18,y=UMPC18]{Data/DetailedCases.dat};
		\addplot[color=Orange,smooth,thick] table [x=X18,y=UOUT18]{Data/DetailedCases.dat};
	\end{axis}
\end{tikzpicture}
\begin{tikzpicture}
	\begin{axis}[height=0.18\textwidth,width=0.32\textwidth,
        legend style={draw=none, fill opacity=0.0, text opacity = 1},legend columns=1, legend pos=south west,
		ylabel=$v_x$ (m/s),xmin=0,xmax=34,ymin=4,ymax=8]	
		\draw[densely dashed] (20,4)--(20,8);
		\addplot[color=Dandelion,smooth,thick] table [x=X18,y=VREF18]{Data/DetailedCases.dat};
		\addplot[color=NavyBlue,smooth,thick] table [x=X18,y=VREAL18]{Data/DetailedCases.dat};
        \legend{\scriptsize ${v_x}\rf$,\scriptsize $\hat{v}_x$};
	\end{axis}
\end{tikzpicture}
\begin{tikzpicture}
	\begin{axis}[height=0.18\textwidth,width=0.32\textwidth,
		ylabel=$y\e$ (m),xmin=0,xmax=34,ymin=-5,ymax=10]	
		\draw[densely dashed] (20,-5)--(20,10);
		\draw[Red,fill=Red,] (19,-2.4) rectangle (21,2.4);
		\addplot[color=Violet,smooth,thick,] table [x=X07,y=Y07]{Data/Experiments.dat};
		\node[right] at (0,7) {\scriptsize $\nu = 0.26$};
		\node[right] at (0,3.5) {\scriptsize $\tau = 3$s};
	\end{axis}
\end{tikzpicture}
\begin{tikzpicture}
	\begin{axis}[height=0.18\textwidth,width=0.32\textwidth,
        legend style={draw=none, fill opacity=0.0, text opacity = 1},legend columns=1, legend pos=north east,
		ylabel=$\delta$ (m/s),xmin=0,xmax=34,ymin=-0.3,ymax=0.3]	
		\draw[densely dashed] (20,-0.3)--(20,0.3);
		\addplot[color=Emerald,smooth,thick] table [x=X07,y=UMPC07]{Data/DetailedCases.dat};
		\addplot[color=Orange,smooth,thick] table [x=X07,y=UOUT07]{Data/DetailedCases.dat};
        \legend{\scriptsize $\delta\mpc$,\scriptsize $\delta$};
	\end{axis}
\end{tikzpicture}
\begin{tikzpicture}
	\begin{axis}[height=0.18\textwidth,width=0.32\textwidth,
		ylabel=$v_x$ (m/s),xmin=0,xmax=34,ymin=4,ymax=8]	
		\draw[densely dashed] (20,4)--(20,8);
		\addplot[color=Dandelion,smooth,thick] table [x=X07,y=VREF07]{Data/DetailedCases.dat};
		\addplot[color=NavyBlue,smooth,thick] table [x=X07,y=VREAL07]{Data/DetailedCases.dat};
	\end{axis}
\end{tikzpicture}
\begin{tikzpicture}
	\begin{axis}[height=0.18\textwidth,width=0.32\textwidth,xlabel=$x\e$ (m),
		ylabel=$y\e$ (m),xmin=0,xmax=34,ymin=-5,ymax=10]	
		\draw[densely dashed] (20,-5)--(20,10);
		\draw[Red,fill=Red,] (19,-3.2) rectangle (21,3.2);
		\addplot[color=Violet,smooth,thick,] table [x=X06,y=Y06]{Data/Experiments.dat};
		\node[right] at (0,7) {\scriptsize $\nu = 0.34$};
		\node[right] at (0,3.5) {\scriptsize $\tau = 3$s};
	\end{axis}
\end{tikzpicture}
\begin{tikzpicture}
	\begin{axis}[height=0.18\textwidth,width=0.32\textwidth,
		xlabel=$x\e$ (m),ylabel=$\delta$ (m/s),xmin=0,xmax=34,ymin=-0.3,ymax=0.3]	
		\draw[densely dashed] (20,-0.3)--(20,0.3);
		\addplot[color=Emerald,smooth,thick] table [x=X06,y=UMPC06]{Data/DetailedCases.dat};
		\addplot[color=Orange,smooth,thick] table [x=X06,y=UOUT06]{Data/DetailedCases.dat};
	\end{axis}
\end{tikzpicture}
\begin{tikzpicture}
	\begin{axis}[height=0.18\textwidth,width=0.32\textwidth,
		xlabel=$x\e$ (m),ylabel=$v_x$ (m/s),xmin=0,xmax=34,ymin=4,ymax=8]	
		\draw[densely dashed] (20,4)--(20,8);
		\addplot[color=Dandelion,smooth,thick] table [x=X06,y=VREF06]{Data/DetailedCases.dat};
		\addplot[color=NavyBlue,smooth,thick] table [x=X06,y=VREAL06]{Data/DetailedCases.dat};
	\end{axis}
\end{tikzpicture}
\caption{Experimental assessment of the $\nu$ influence: vehicle trajectory for the desired velocity $v_x = 6$m/s.}\label{fig:exptx6}
\end{figure*}
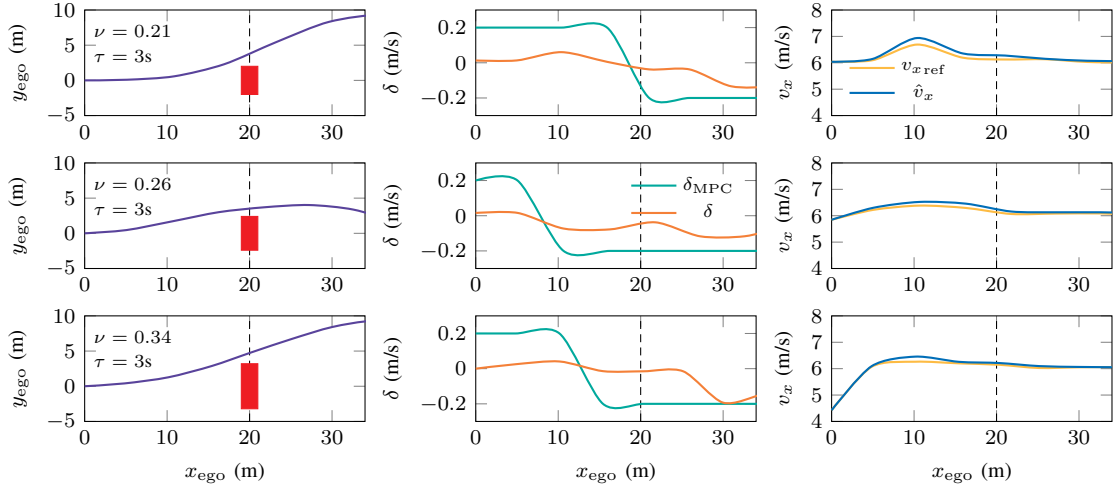

Three examples for the next set of experiments are shown in Fig.~\ref{fig:exptx6}. The distance to the static obstacle at the detection time is set to $\tau = 3$s for all three cases, while varying the value of $\nu$ from $0.4$ to $0.35$. Similar to the relaxed cases, we observe that MPC is still ``overreacting'' due to the limited computation time, while the combined planner manages to avoid collision without an extreme steering command. the desired velocity is set to $6$m/s which is the same as the longitudinal velocity at the time of obstacle detection in the first two cases. However, in the third case, the vehicle detects the obstacle at the speed of $4.5$m/s which means that a much higher steering command is required to achieve the same lateral displacement. As a result, we observe that the planners opt for an increase in the longitudinal velocity to allow for a lower and safer steering command. 
\begin{obsrv}
    With a higher velocity, we observe that the GPS measurement lag can be more limiting in a swift return to the initial lateral position after overtaking the obstacle.
\end{obsrv}

\subsection{Extreme cases with $v_x = 7$m/s}

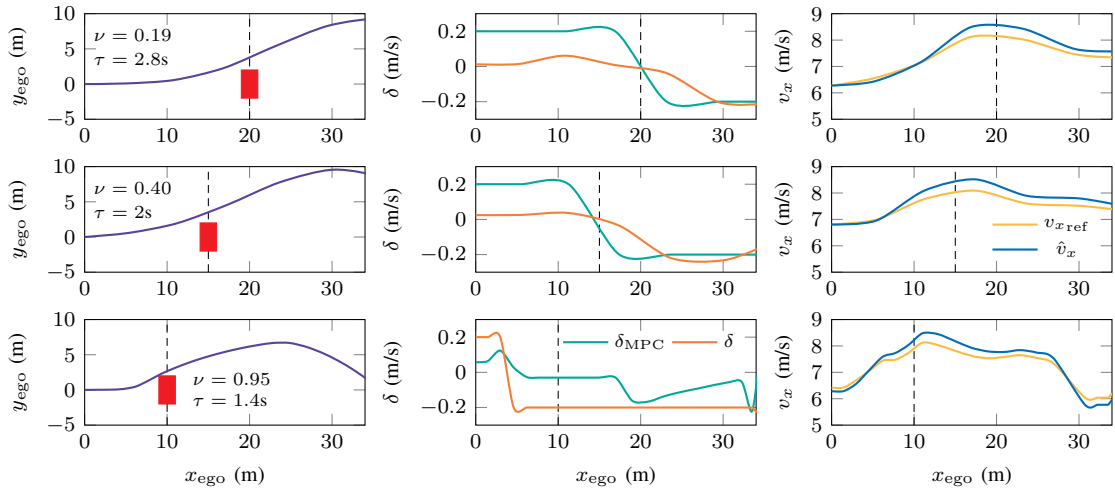
\begin{figure*}[hbtp]\centering
\begin{tikzpicture}
	\begin{axis}[height=0.18\textwidth,width=0.32\textwidth,
		ylabel=$y\e$ (m),xmin=0,xmax=34,ymin=-5,ymax=10]	
		\draw[densely dashed] (20,-5)--(20,10);
		\draw[Red,fill=Red,] (19,-2.0) rectangle (21,2.0);
		\addplot[color=Violet,smooth,thick,] table [x=X10,y=Y10]{Data/Experiments.dat};
		\node[right] at (0,7) {\scriptsize $\nu = 0.19$};
		\node[right] at (0,3.5) {\scriptsize $\tau = 2.8$s};
	\end{axis}
\end{tikzpicture}
\begin{tikzpicture}
	\begin{axis}[height=0.18\textwidth,width=0.32\textwidth,
		ylabel=$\delta$ (m/s),xmin=0,xmax=34,ymin=-0.3,ymax=0.3]	
		\draw[densely dashed] (20,-0.3)--(20,0.3);
		\addplot[color=Emerald,smooth,thick] table [x=X10,y=UMPC10]{Data/DetailedCases.dat};
		\addplot[color=Orange,smooth,thick] table [x=X10,y=UOUT10]{Data/DetailedCases.dat};
	\end{axis}
\end{tikzpicture}
\begin{tikzpicture}
	\begin{axis}[height=0.18\textwidth,width=0.32\textwidth,
		ylabel=$v_x$ (m/s),xmin=0,xmax=34,ymin=5,ymax=9]	
		\draw[densely dashed] (20,5)--(20,9);
		\addplot[color=Dandelion,smooth,thick] table [x=X10,y=VREF10]{Data/DetailedCases.dat};
		\addplot[color=NavyBlue,smooth,thick] table [x=X10,y=VREAL10]{Data/DetailedCases.dat};
	\end{axis}
\end{tikzpicture}
\begin{tikzpicture}
	\begin{axis}[height=0.18\textwidth,width=0.32\textwidth,
		ylabel=$y\e$ (m),xmin=0,xmax=34,ymin=-5,ymax=10]	
		\draw[densely dashed] (15,-5)--(15,10);
		\draw[Red,fill=Red,] (14,-2.0) rectangle (16,2.0);
		\addplot[color=Violet,smooth,thick,] table [x=X15,y=Y15]{Data/Experiments.dat};
		\node[right] at (0,7) {\scriptsize $\nu = 0.40$};
		\node[right] at (0,3.5) {\scriptsize $\tau = 2$s};
	\end{axis}
\end{tikzpicture}
\begin{tikzpicture}
	\begin{axis}[height=0.18\textwidth,width=0.32\textwidth,
		ylabel=$\delta$ (m/s),xmin=0,xmax=34,ymin=-0.3,ymax=0.3]	
		\draw[densely dashed] (15,-0.3)--(15,0.3);
		\addplot[color=Emerald,smooth,thick] table [x=X15,y=UMPC15]{Data/DetailedCases.dat};
		\addplot[color=Orange,smooth,thick] table [x=X15,y=UOUT15]{Data/DetailedCases.dat};
	\end{axis}
\end{tikzpicture}
\begin{tikzpicture}
	\begin{axis}[height=0.18\textwidth,width=0.32\textwidth,
        legend style={draw=none, fill opacity=0.0, text opacity = 1},legend columns=1, legend pos=south east,
		ylabel=$v_x$ (m/s),xmin=0,xmax=34,ymin=5,ymax=9]	
		\draw[densely dashed] (15,5)--(15,9);
		\addplot[color=Dandelion,smooth,thick] table [x=X15,y=VREF15]{Data/DetailedCases.dat};
		\addplot[color=NavyBlue,smooth,thick] table [x=X15,y=VREAL15]{Data/DetailedCases.dat};
        \legend{\scriptsize ${v_x}\rf$,\scriptsize $\hat{v}_x$};
	\end{axis}
\end{tikzpicture}
\begin{tikzpicture}
	\begin{axis}[height=0.18\textwidth,width=0.32\textwidth,
		xlabel=$x\e$ (m),ylabel=$y\e$ (m),xmin=0,xmax=34,ymin=-5,ymax=10]	
		\draw[densely dashed] (10,-5)--(10,10);
		\draw[Red,fill=Red,] (9,-2.0) rectangle (11,2.0);
		\addplot[color=Violet,smooth,thick,] table [x=X08,y=Y08]{Data/Experiments.dat};
		\node[right] at (12,1.5) {\scriptsize $\nu = 0.95$};
		\node[right] at (12,-1.5) {\scriptsize $\tau = 1.4$s};
	\end{axis}
\end{tikzpicture}
\begin{tikzpicture}
	\begin{axis}[height=0.18\textwidth,width=0.32\textwidth,
        legend style={draw=none, fill opacity=0.0, text opacity = 1},legend columns=2, legend pos=north east,
		xlabel=$x\e$ (m),ylabel=$\delta$ (m/s),xmin=0,xmax=34,ymin=-0.3,ymax=0.3]	
		\draw[densely dashed] (10,-0.3)--(10,0.3);
		\addplot[color=Emerald,smooth,thick] table [x=X08,y=UMPC08]{Data/DetailedCases.dat};
		\addplot[color=Orange,smooth,thick] table [x=X08,y=UOUT08]{Data/DetailedCases.dat};
        \legend{\scriptsize $\delta\mpc$,\scriptsize $\delta$};
	\end{axis}
\end{tikzpicture}
\begin{tikzpicture}
	\begin{axis}[height=0.18\textwidth,width=0.32\textwidth,
		xlabel=$x\e$ (m),ylabel=$v_x$ (m/s),xmin=0,xmax=34,ymin=5,ymax=9]	
		\draw[densely dashed] (10,5)--(10,9);
		\addplot[color=Dandelion,smooth,thick] table [x=X08,y=VREF08]{Data/DetailedCases.dat};
		\addplot[color=NavyBlue,smooth,thick] table [x=X08,y=VREAL08]{Data/DetailedCases.dat};
	\end{axis}
\end{tikzpicture}
\caption{Experimental assessment of the hazard: vehicle trajectory for the desired velocity $v_x = 7$m/s.}\label{fig:exptx7}
\end{figure*}

Finally, we conduct the extreme cases close to the maximum allowed longitudinal velocity on campus field with setting the desired vehicle speed to $7$m/s. Figure~\ref{fig:exptx7} shows the results for three cases with the same obstacle size, while changing the $\tau$ and $\nu$ values. In all three cases, we observe that the planners need to increase the longitudinal velocity as the steering limits are not sufficient to guarantee a safe collision-free maneuver. 
\begin{obsrv}
    Increasing the longitudinal velocity is a potential solution in scenarios where the steering limits are not sufficient to guarantee a safe collision-free maneuver.
\end{obsrv}
In the most extreme case with $\tau = 1.4$s and $\nu=0.95$, we observe that the MPC cannot find a feasible solution to avoid colliding with the static obstacle. However, the feed-forward planner offers a solution close to the limits of steering which helps in avoiding the highly-probable collision. While returning to the initial lateral position and the desired velocity after overtaking the obstacle.

\subsection{Comparison with MPC}

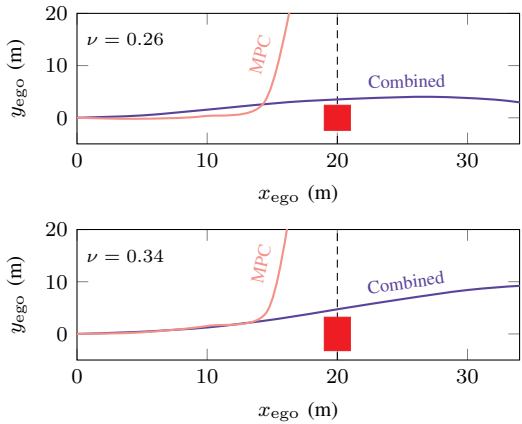
\begin{figure}[hbtp]\centering
\begin{tikzpicture}
	\begin{axis}[height=0.2\textwidth,width=0.45\textwidth,xlabel=$x\e$ (m),
		ylabel=$y\e$ (m),xmin=0,xmax=34,ymin=-5,ymax=20]	
		\draw[densely dashed] (20,-5)--(20,20);
		\draw[Red,fill=Red,] (19,-2.4) rectangle (21,2.4);
		\addplot[color=Violet,smooth,thick,] table [x=X07,y=Y07]{Data/Experiments.dat}
		node[midway,sloped,above,xshift=-3pt] {\scriptsize Combined};
		\addplot[color=Salmon,smooth,thick,] table [x=OX1,y=OY1]{Data/Experiments.dat} node[midway,sloped,above,xshift=10pt] {\scriptsize MPC};
		\node[right] at (0,15) {\scriptsize $\nu = 0.26$};
	\end{axis}
\end{tikzpicture}
\begin{tikzpicture}
	\begin{axis}[height=0.2\textwidth,width=0.45\textwidth,xlabel=$x\e$ (m),
		ylabel=$y\e$ (m),xmin=0,xmax=34,ymin=-5,ymax=20]	
		\draw[densely dashed] (20,-5)--(20,20);
		\draw[Red,fill=Red,] (19,-3.2) rectangle (21,3.2);
		\addplot[color=Violet,smooth,thick,] table [x=X06,y=Y06]{Data/Experiments.dat}
		node[midway,sloped,above,xshift=-3pt] {\scriptsize Combined};
		\addplot[color=Salmon,smooth,thick,] table [x=OX2,y=OY2]{Data/Experiments.dat}
		node[midway,sloped,above,xshift=10pt] {\scriptsize MPC};
		\node[right] at (0,15) {\scriptsize $\nu = 0.34$};
	\end{axis}
\end{tikzpicture}
\caption{Comparing the MPC planner in~\cite{Ammour2022} and our proposed combined planning strategy}\label{fig:complit}
\end{figure}

In the final set of tests, we compare the collision avoidance capabilities of the MPC planner against our proposed combined strategy. Figure~\ref{fig:complit} shows two scenarios with $v_x = 6$m/s and $\tau = 3$s with two different obstacle widths. It can be observed that while MPC overreacts to the presence of the obstacle by providing extreme steering commands due to limited computation time, the combined planner can safely avoid colliding with the obstacle while keeping the ego vehicle in the field limits.

\section{Conclusions}\label{sec:conclude}

This paper has offered experimental insights into real-time implementation of MPC for collision avoidance after the unexpected appearance of a static obstacle. Given the limitations of state-of-the-art nonlinear MPC in providing feasible solutions in real-time, we have proposed a human-inspired feed-forward planner to support situations where the MPC optimization problem is either infeasible or converges to a poor local solution due to a poor initial guess. Our real-world experiments, conducted under various conditions and speeds using the FPEV2-Kanon electric vehicle, validate the effectiveness of our proposed planning strategy, also in comparison to a state-of-the-art MPC motion planner.

For future research, we suggest real-time experimental tests considering parametric uncertainties e.g.~due to variations in friction coefficient.

\section*{Acknowledgment}

This research was supported by the Dutch Research Council (NWO-TTW) within the project EVOLVE (nr.~18484), the European Union Horizon 2020 innovation and research program within the Marie Skłodowska-Curie project OWHEEL (nr.~872907), the Industrial Technology Research Grant Program from New Energy and Industrial Technology Development Organization (NEDO) of Japan under Grant 05A48701d, and the Ministry of Education, Culture, Sports, Science, and Technology under Grant 22246057, Grant 26249061, and Grant 22K14283.

\ifCLASSOPTIONcaptionsoff
  \newpage
\fi

\bibliographystyle{IEEEtran}
\bibliography{Citations}

%

\vspace{-10mm}

\begin{IEEEbiography}[{\includegraphics[width=1in,height=1.25in,clip,keepaspectratio]{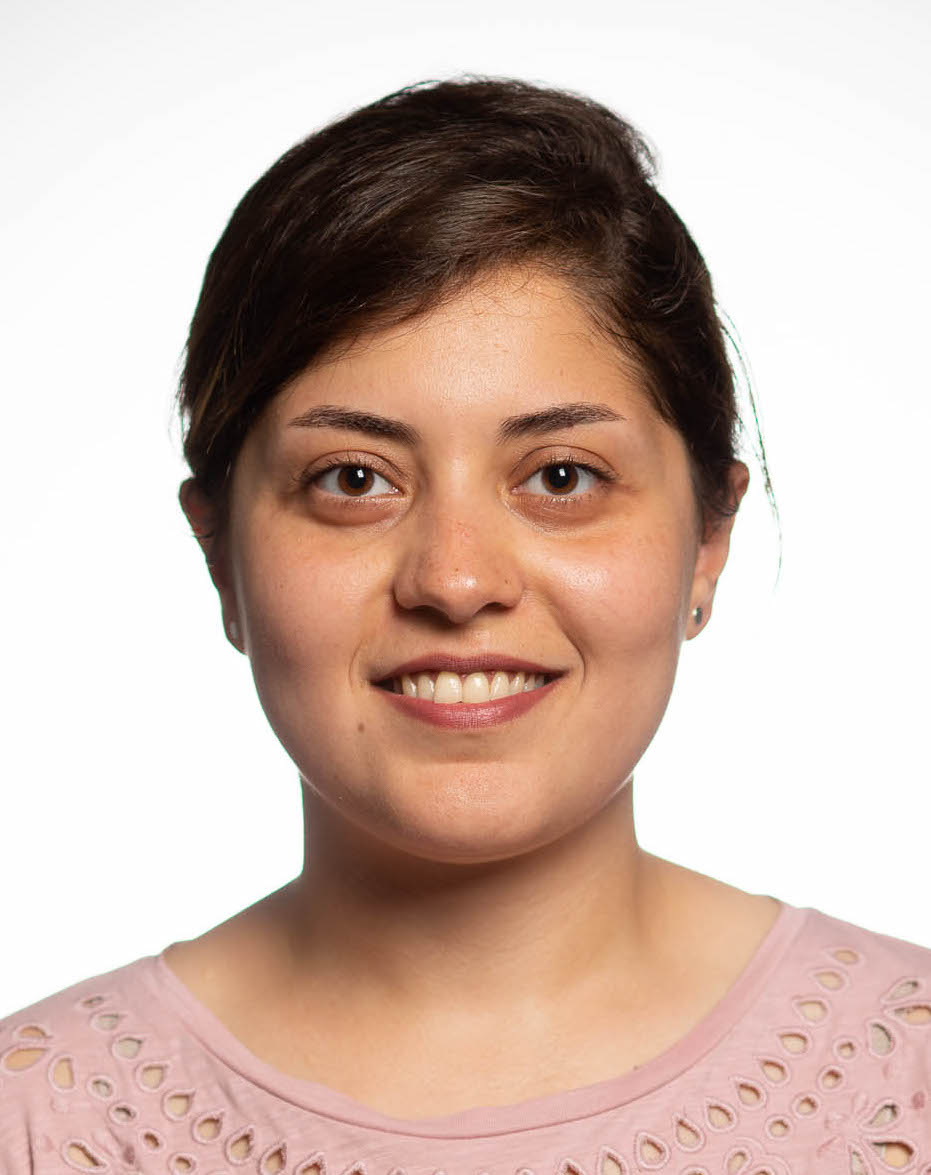}}]{Leila Gharavi} obtained her PhD in systems and control in 2025 from Delft University of Technology, The Netherlands. She received her BSc and MSc degrees in mechanical engineering from Amirkabir University of Technology (Tehran Polytechnic) in Iran and has research experience in automatic production, vibration analysis and control of nonlinear dynamics, and soft rehabilitation robotics. 
\end{IEEEbiography}
\vspace{-10mm}
\begin{IEEEbiography}[{\includegraphics[width=1in,height=1.25in,clip,keepaspectratio]{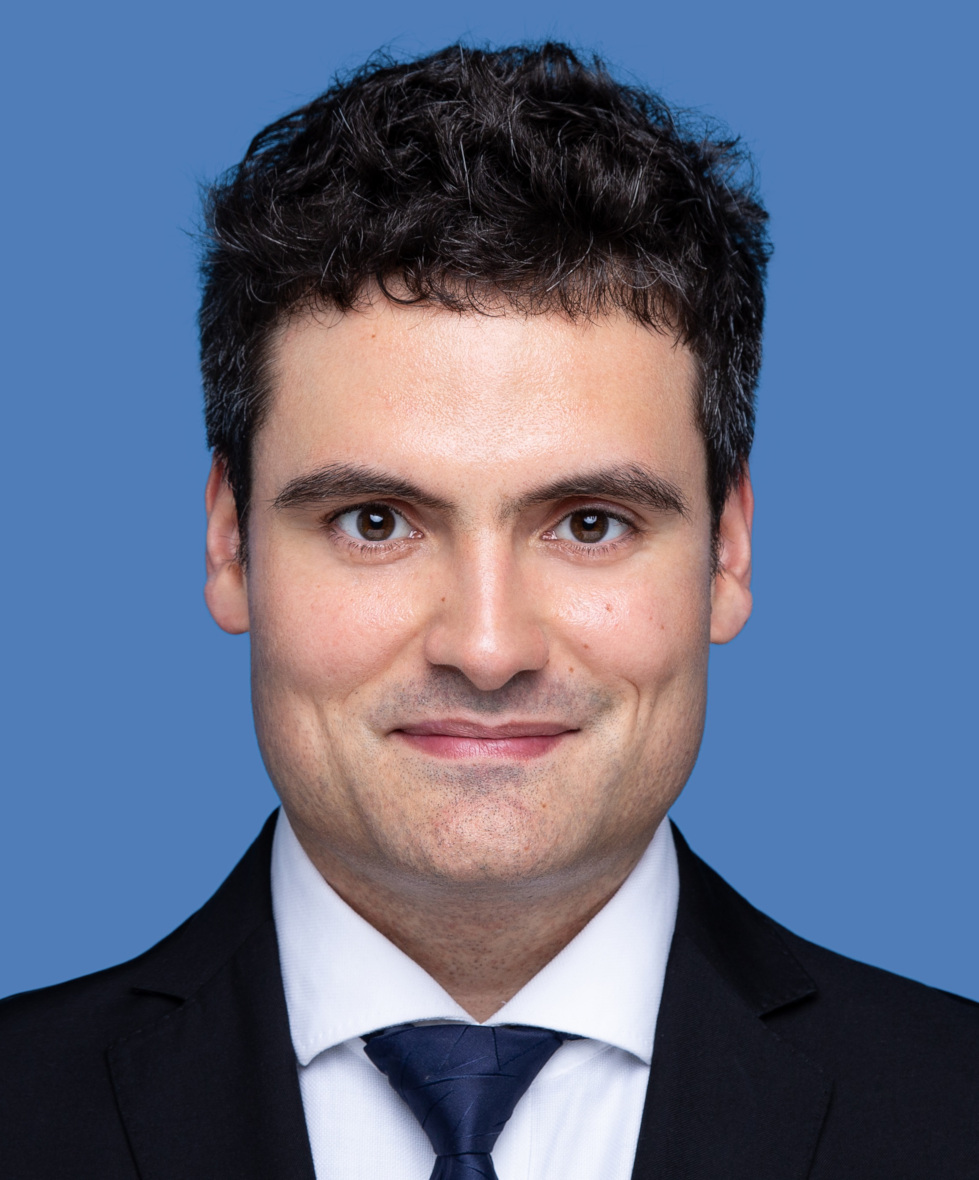}}]{Simone Baldi} 
received the B.Sc. degree in electrical engineering, and the M.Sc. and Ph.D. degrees in automatic control systems engineering from the University of Florence, Florence, Italy, in 2005, 2007, and 2011, respectively. He is a Professor with the School of Mathematics, Southeast University, China. His research interests include adaptive and learning systems with applications in smart energy and intelligent vehicle systems.
\end{IEEEbiography}
\vspace{-10mm}
\begin{IEEEbiography}[{\includegraphics[width=1in,height=1.25in,clip,keepaspectratio]{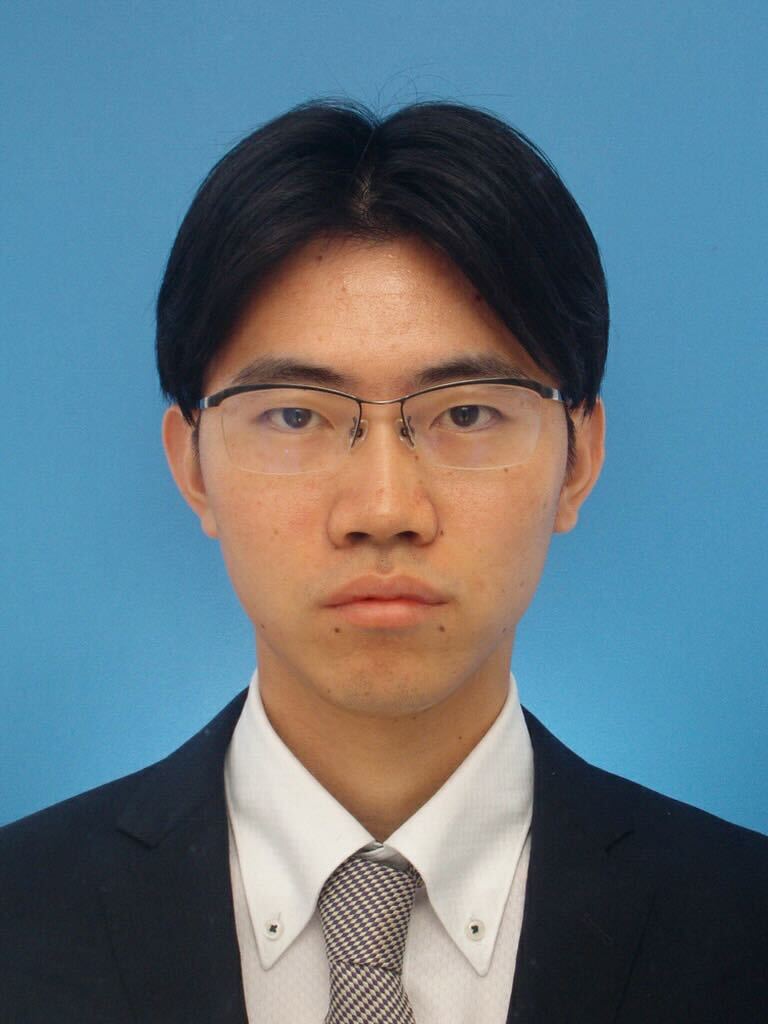}}]{Yuki Hosomi} is currently pursuing the M.S. degree with the Department of Advanced Energy, Graduate School of Frontier Sciences, The University of Tokyo. His research interests includes motion control and energy optimization of electric vehicles.

\end{IEEEbiography}
\vspace{-10mm}
\begin{IEEEbiography}[{\includegraphics[width=1in,height=1.25in,clip,keepaspectratio]{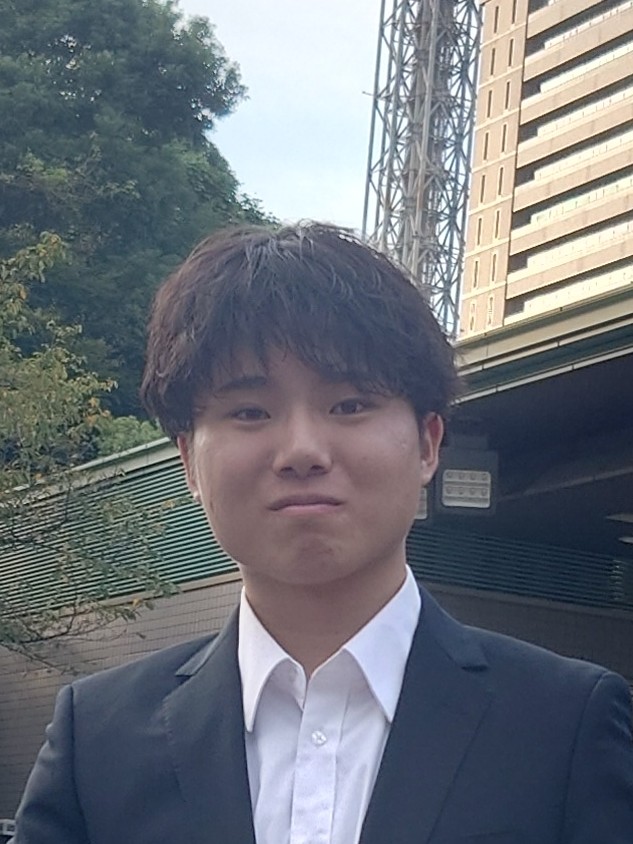}}]{Sato Tona} is currently pursuing the M.S. degree with the Department of Advanced Energy, Graduate School of Frontier Sciences, The University of Tokyo. His research interests includes motion control of electric vehicles.

\end{IEEEbiography}
\vspace{-10mm}
\begin{IEEEbiography}[{\includegraphics[width=1in,height=1.25in,clip,keepaspectratio]{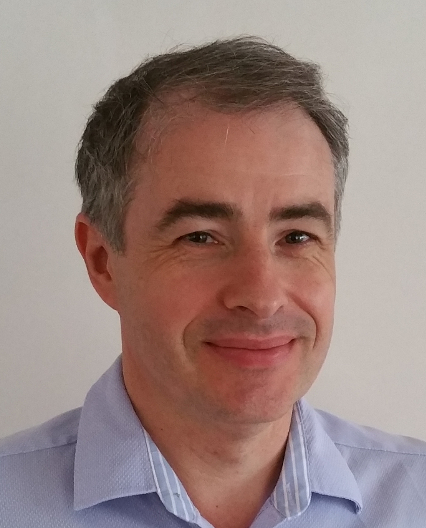}}]{Bart De Schutter} 
is a Full Professor with the Delft Center for Systems and Control, Delft University of Technology, Delft, The Netherlands. He is a Senior Editor of the IEEE Transactions on Intelligent Transportation Systems. His current research interests include intelligent transportation, infrastructure and power systems, hybrid systems, and multi-level control.
\end{IEEEbiography}
\vspace{-10mm}
\begin{IEEEbiography}[{\includegraphics[width=1in,height=1.25in,clip,keepaspectratio]{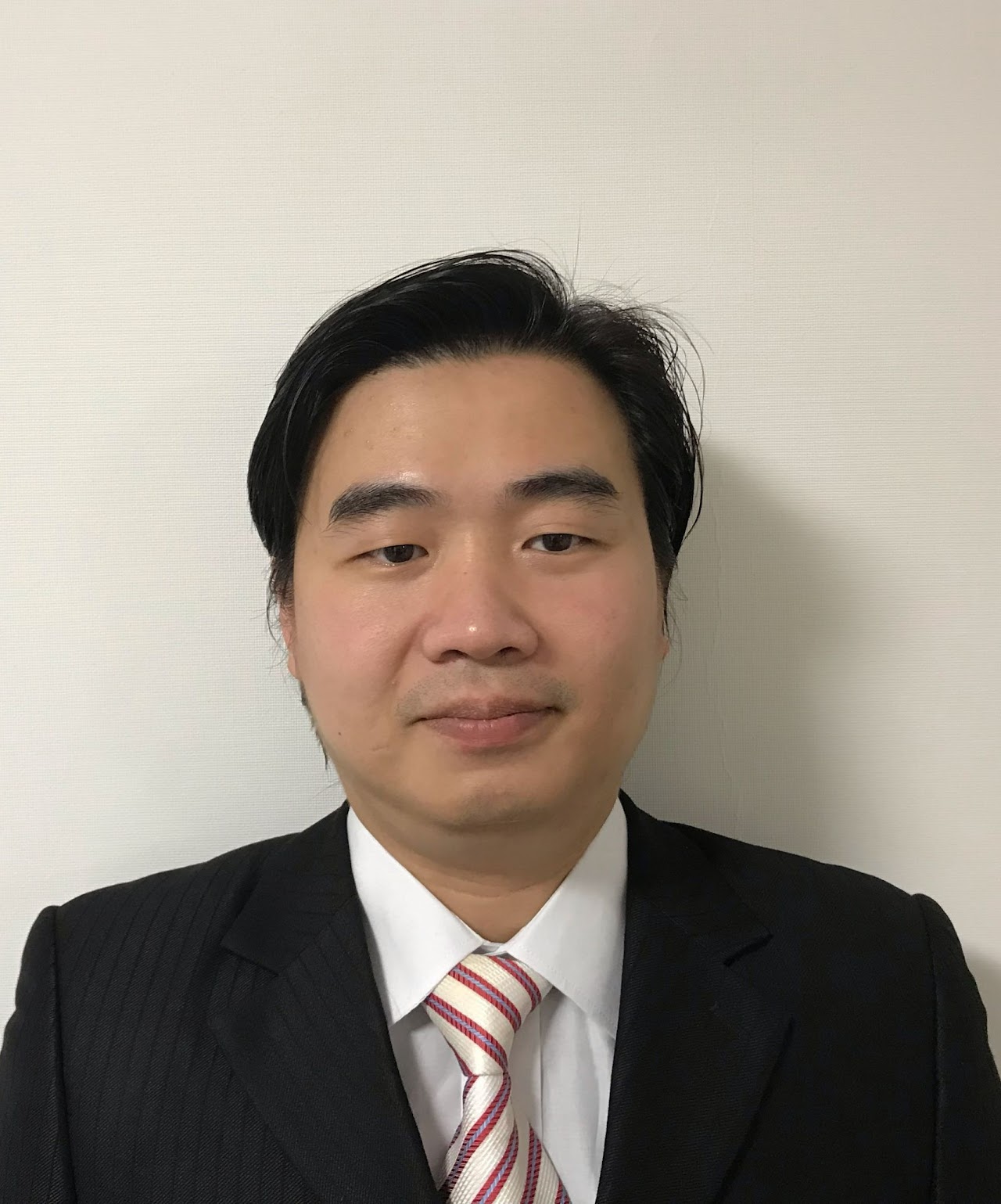}}]{Binh-Minh Nguyen} is an Assistanct Professor with the Department of Advanced Energy, the University of Tokyo. His main research interests include glocal control, robust control, and their application to electric vehicles, flying cehilces, and power systems.


\end{IEEEbiography}

\vspace{-10mm}
\begin{IEEEbiography}[{\includegraphics[width=1in,height=1.25in,clip,keepaspectratio]{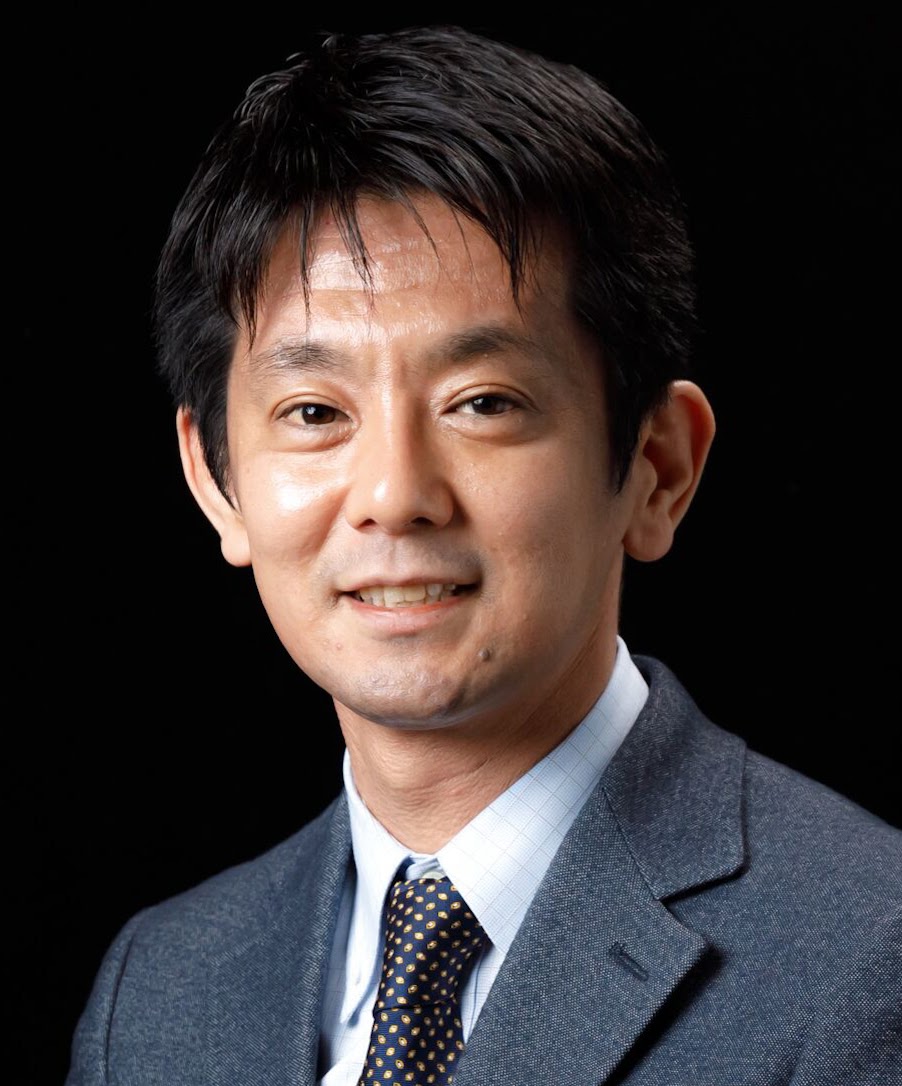}}]{Hiroshi Fujimoto} is a Professor with the Department of Advanced Energy, the University of Tokyo. His main research interests include control engineering, motion control, nano-scale servo systems, electric vehicle, motor drive, visual servoing, and wireless power transfer. He is a Fellow Member of the IEEE and a Senior Member of the IEE of Japan.
\end{IEEEbiography}







\end{document}